%% file: a1main.tex
\documentclass[preprint,12pt, sort&compress]{elsarticle}

\usepackage{amsmath}

\graphicspath{{pics/}}

\usepackage{multirow}
\usepackage{rotating}
\usepackage{makecell}
\usepackage{color,soul}
\usepackage{siunitx}

\journal{Nuclear Instruments and Methods A}

\begin{document}

\begin{frontmatter}



\title{A new method to find out the optimal neutron moderator size based on neutron scattering instrument parameters}


\author[FZJ]{Petr Konik}

\affiliation[FZJ]{organization={Jülich Centre for Neutron Science (JCNS) at Heinz Maier-Leibnitz Zentrum (MLZ), Forschungszentrum Jülich GmbH},
            addressline={Lichtenbergstr.~1}, 
            city={Garching},
            postcode={85748},
            country={Germany}}

\author[FZJ]{Alexander Ioffe}

\begin{abstract}

Simple and fast analytic method to find optimal combinations of sizes of neutron moderator and optical system entrance, allowing for the full sample illumination with minimum to none background, is developed. In the case of employing low-dimensional para-hydrogen moderators with enhanced neutron beam brilliance, the method allows to determine the minimum size of moderator, which provides the highest sample flux while keeping the sample fully illuminated. 

This method can be used during the design of new neutron sources, upgrades of neutron optical systems and moderator replacements.
\end{abstract}



\end{frontmatter}



\input{intro.tex}

\input{notations.tex}

\input{requirements.tex}

\input{guide_exit.tex}

\input{guide_entrance.tex}

\input{COFSI.tex}

\input{sample_flux_analytic.tex}

\input{sample_flux_MC.tex}

\input{para_hydrogen.tex}

\input{two_modes.tex}

\input{TOF.tex}

\input{two_instruments.tex}

\input{outro.tex}

\section*{Acknowledgement}

This project has received funding from the European Union's Horizon 2020 research and innovation programme under grant agreement No. 871072.

\bibliographystyle{elsarticle-num} 
\bibliography{cas-refs}

\end{document}

%% file: intro.tex
\section{Introduction}


In the last decade low-dimensional neutron moderators filled with almost pure para-hydrogen have been developed~\cite{batkov2013,mezei2014}. Thanks to the large difference between scattering cross-sections for thermal and cold neutrons, such tube-like or disk-like moderators can provide significantly higher brilliance than traditional voluminous cold moderators. Investigations carried out at ESS showed potential gains of 2--3 times in neutron brilliance when reducing the moderator size from 12 cm to 3 cm~\cite{batkov2013}. However, in many cases the use of such small moderator results in the under-illumination of the neutron transport system (NTS), i.e. moderator does not provide all neutrons with trajectories, which could be accepted by NTS. Indeed, the full illumination of the NTS entrance and the sample is required to reach the maximum sample flux.

However, if the neutron beam of a too large divergence or a too large size is delivered to the sample, then the over-illumination occurs, that is undesirable since such “useless” neutrons may lead to increase of background and thus worsening experimental conditions. Hence, neutron scattering instrument requirements for sample size and angular divergence impose constraints on the parameters of the neutron beam leaving the NTS.




Therefore, the question arises, which combinations of moderator size and the entrance size of NTS allow to obtain optimal sample illumination. Until the present the straightforward approach to this problem has been used, involving extensive and time-consuming Monte-Carlo simulations of both, moderator and neutron optics~\cite{andersen2018, ma2021}. Few attempts have been made to constrain the optimization process by reducing the number of free optimization parameters~\cite{zhao2013, bertelsen2016}.

In this article we describe the analytic method allowing to quickly find combinations of moderator and NTS entrance sizes that are optimal in the sense of providing maximum sample flux with minimum background. While not fully eliminating the need of Monte-Carlo simulations for the neutron optics optimization, this method considerably decreases the number of free parameters and allows to decouple the optimizations of moderator and neutron optics.

The rest of the paper is structured as follows.

In Sec.~\ref{notations} the used notations and the phase space terminology are introduced.

In Sec.~\ref{PS_considerations} we use phase space formalism to investigate connections between the neutron scattering instrument requirements, NTS properties and moderator size. The result is the simple expression allowing to calculate optimal moderator size for any given instrument parameters.

In Sec.~\ref{sample_flux} we investigate the influence of deviations from the optimal moderator size on the sample flux. Special attention is given to the para-hydrogen moderators with size-dependent brilliance.

Finally, in Sec.~\ref{practice} we give examples of practical applications of the developed method. The cases of varying instrument parameters according to experimental needs, as well as the optimization of a single moderator serving multiple instruments are explored.

%% file: notations.tex
\section{Notations}
\label{notations}

\begin{figure}[b!]
\begin{center}
\begin{minipage}[h]{0.99\linewidth}
\includegraphics[width=1\linewidth]{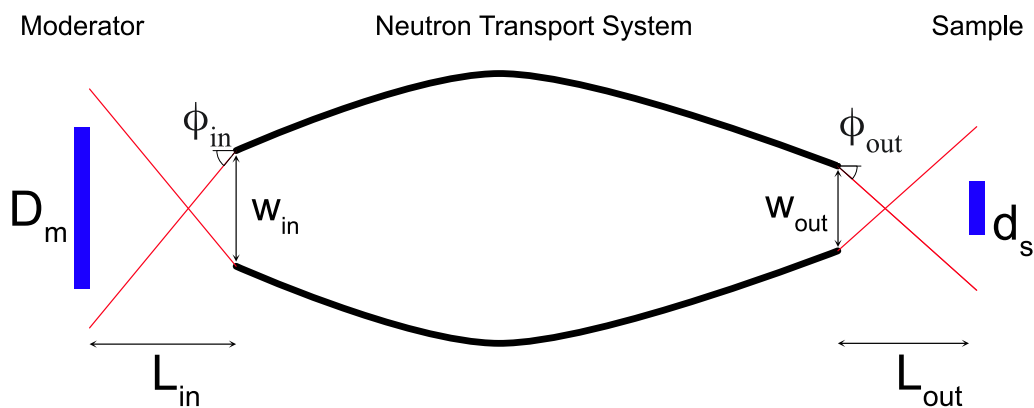}
\end{minipage}

\caption{Layout of a neutron instrument from the optical point of view.}
\label{guide_scheme}
\end{center}
\end{figure}

Since our aim is to investigate the sample illumination conditions (under- or over-illumination), we consider neutron instrument holistically starting with the neutron moderator, ending with the sample and disregarding the analyser-detector part. We call the set of all elements situated between the moderator and sample the neutron transport system (NTS), which may include neutron guides~\cite{christ1962, mezei2000, schanzer2004}, focusing mirrors~\cite{kentzinger2004, stahn2011, pipich2015}, nested optics~\cite{mildner2011, khaykovich2011, herb2022}, lenses~\cite{choi2000, oku2004, fuzi2004}, monochromators, entrance and exit slits, Soller collimators, etc. The general layout of neutron instrument is shown in Fig.~\ref{guide_scheme}. Here $D_m$ is the source size, $d_s$ is the sample size, $L_{in}$ is the distance between the source and the NTS entrance, and $L_{out}$ is the distance from the NTS exit to the sample. NTS entrance and exit sizes are noted as $w_{in}$ and $w_{out}$, respectively. Finally, $\phi_{in}$ is the neutron beam divergence that can be accepted by NTS, $\phi_{out}$ is the neutron beam divergence at the NTS exit.

For a newly designed instrument only $d_s$, $L_{in}$ and $L_{out}$ are constrained based on instrument's science case, potential radiation or thermal damage to neutron optics and requirements for sample environment size or angular resolution, respectively. All other parameters are free to vary: moderator size $D_m$ has to be adapted to instrument's needs; guide shape, coating, $w_{in}$ and $w_{out}$ have to provide the highest NTS performance, etc.

To evaluate performance of NTS the phase space diagram technique can be employed~\cite{copley1993, fuzi2006, bertelsen2016}. The phase space (PS) is a space, where each point corresponds to the unique state of the system~--- a single neutron trajectory in the neutron beam in our case. This 5-dimensional PS is built upon two space coordinates (i.e. the beam cross-section), two angles defining the direction of beam propagation, and wavelength. The ensemble of all points (neutron trajectories) in PS is called the phase space volume $V$ and constitutes the whole neutron beam.

The brilliance $b$ of the neutron beam is the the PS density defined as the total number $\Phi$ of neutron trajectories divided by the PS volume they occupy: 

\begin{equation}
    b = \dfrac{\Phi}{V}.
    \label{brilliance}
\end{equation}

Usually the neutron wavelength is not changed during the beam propagation through NTS, allowing to exclude this coordinate from further considerations. Assuming rectangular cross-section of NTS, we can study separately two PS projections, horizontal and vertical, each built with one space and one angular coordinates. From here on we use the term “phase space” referring to one of those projections. Phase space volumes corresponding to each of these projection are 2-dimensional, so that brilliance is measured in $\dfrac{\text{n}}{\text{s} \cdot \text{cm} \cdot \text{rad}}$, rather than in $\dfrac{\text{n}}{\text{s} \cdot \text{cm}^2 \cdot \text{sr}}$ as usual. Note, the method derived in this paper can be easily expanded to account for the 5-dimensional PS as a whole, however corresponding formulas become cumbersome.

According to Liouville theorem brilliance $b_{out}$ at the NTS exit cannot be larger than brilliance $b_{in}$ at its entrance :

\begin{equation}
    b_{out} \le b_{in}.
\end{equation}
For optical system with no transmission losses (ideal brilliance transfer) $b_{out} = b_{in}$ and

\begin{equation}
    V_{out} = V_{in},
    \label{ideal_trans}
\end{equation}
where $V_{out}$ and $V_{in}$ are the PS volumes of the beam at the NTS exit and entrance, respectively. We use this assumption for the rest of the paper, except for Sec.~\ref{sample_flux_mc}, where transmission losses are taken into account.

%% file: requirements.tex
\newpage
\section{Choice of optimal sizes of moderator and NTS entrance}
\label{PS_considerations}

\subsection{Instrument requirements}

To achieve the expected instrument performance each point of the sample with size $d_s$ must be illuminated by a neutron beam with divergence $2\alpha_s$, where $\alpha_s$ is required geometric resolution defined by momentum transfer or energy resolution.

\begin{figure}[h!]
\begin{center}
\begin{minipage}[h]{0.9\linewidth}
\includegraphics[width=1\linewidth]{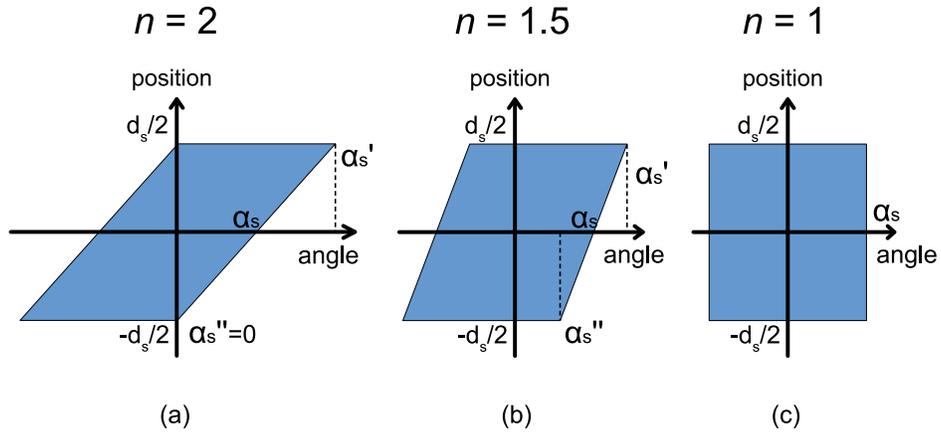}
\end{minipage}

\caption{Different PS volumes $V_s$, which can be required by instrument.}
\label{parallelogram}
\end{center}
\end{figure}

Corresponding PS volume $V_s$ has a shape of parallelogram and is equal to

\begin{equation}
    V_s = 2 d_s \alpha_s = d_s (\alpha_s' + \alpha_s''),
    \label{sample_volume}
\end{equation}

\begin{equation}
    \alpha_s = \dfrac{\alpha_s' + \alpha_s''}{2},
    \label{angles}
\end{equation}
where $\alpha_s'$ and $\alpha_s''$ define positions of parallelogram corners (Fig.~\ref{parallelogram}).

Here we introduce factor $n\ge1$ describing the inclination of the parallelogram:

\begin{equation}
    \alpha_s' = n \alpha_s.
    \label{factor}
\end{equation}
Taking into account~(\ref{angles}) we can write
\begin{equation}
    \alpha_s'' = (2 - n) \alpha_s.
    \label{factor2}
\end{equation}

\begin{figure}[t]
\begin{center}
\begin{minipage}[h]{0.99\linewidth}
\includegraphics[width=1\linewidth]{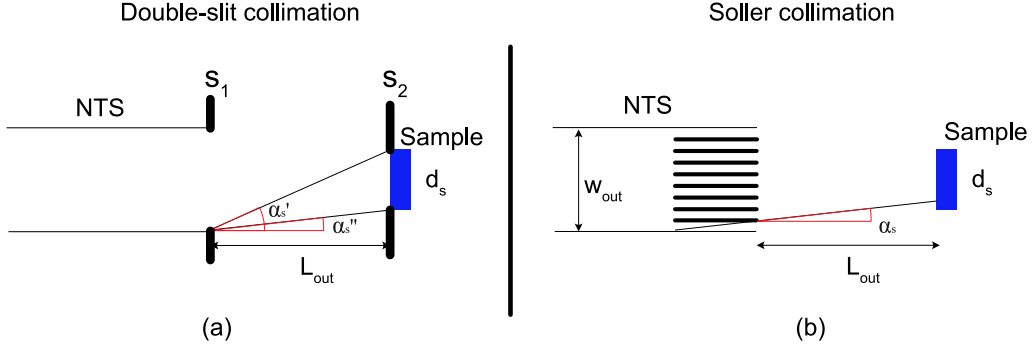}
\end{minipage}

\caption{ (a) Instrument with double-slit collimation. (b) Instrument with Soller collimation.}
\label{2slit}
\end{center}
\end{figure}

Practically $\alpha_s$ is defined by the collimation system of the instrument. In the case of the double-slit collimation, the first slit $s_1$ is the exit aperture of  NTS, so that $s_1 = w_{out}$. If the second slit is placed at the sample $s_2 = d_s$, then $\alpha_s'$ and $\alpha_s''$ correspond to the angles shown in Fig.~\ref{2slit}a and can be calculated as follows:

\begin{equation}
    \alpha_s' = \dfrac{s_1 + s_2}{2 L_{out}},
    \label{slit_equ}
\end{equation}
\begin{equation}
    \alpha_s'' = \dfrac{s_1 - s_2}{2 L_{out}}.
\end{equation}
From here on all equations are given in the small angle approximation, which is usually appropriate for neutron optics. Note that $\alpha_s''$ can be negative, corresponding to the case of the sample being larger than the exit of NTS.

The case of double-slit collimation demonstrates the physical meaning of $n$ as an indicator of the relative size of collimation slits. If the first slit is $x$ times larger than the second one, then 

\begin{equation}
    \alpha_s' = \dfrac{(x+1)s_2}{2L_{out}},
\end{equation}
\begin{equation}
    \alpha_s'' = \dfrac{(x-1)s_2}{2L_{out}},
\end{equation}
\begin{equation}
    \alpha_s = \dfrac{(x+1)s_2 + (x-1)s_2}{4L_{out}} = \dfrac{xs_2}{2L_{out}}.
\end{equation}
Taking into account Eq.~\eqref{factor} we get

\begin{equation}
    n(x) = \dfrac{x + 1}{x}.
\end{equation}
In reflectometry slits of equal sizes are mostly used, meaning $x=1$ and $\alpha_s'' = 0$, what corresponds to $n(1)=2$ (see Fig.~\ref{parallelogram}a). For small-angle scattering experiments $x=2$ is considered to be optimal what corresponds to $n(2)=1.5$ (Fig.~\ref{parallelogram}b).

In case of double-slit collimation $L_{out}$ is the distance between the slits and is not constrained by instrument requirements. Instead, each particular scattering technique requires a certain value $n$. Based on geometric resolution requirement $\alpha_s$, one can calculate collimation base using Eqs.~(\ref{factor},\ref{slit_equ}):

\begin{equation}
    \alpha_s' = n \alpha_s = \dfrac{x+1}{x} \alpha_s,
\end{equation}

\begin{equation}
    \alpha_s' = \dfrac{s_1 + s_2}{2 L_{out}} = \dfrac{(x+1) d_s}{2 L_{out}},
\end{equation}

\begin{equation}
    L_{out} = \dfrac{d_s}{2\alpha_s (n-1)}.
    \label{db-coll}
\end{equation}

Note that in some cases it is not possible to place the second collimation slit directly at the sample. If that happens, in the following considerations we will consider this slit as the “sample”, since optimally the sample should accept all neutrons going through the second slit. The size of this slit should be chosen to provide the full illumination of the real sample.

An alternative way to collimate the neutron beam is to use Soller collimator (Fig.~\ref{2slit}b). In this case divergence at any point of the sample is equal to the collimation angle and PS volume $V_s$ has the rectangular shape: $\alpha_s'' = \alpha_s' = \alpha_s$ and $n=1$ (see Fig.~\ref{parallelogram}c). $L_{out}$ is constrained by geometrical restrictions around the sample, e.g. the size of bulky sample environment.

If neutron instrument does not use any collimation device, then geometric resolution is defined by the natural divergence of neutron beam leaving the NTS (e.g. determined by the critical angle of mirror coating). Since Soller collimator is an integral part of NTS, there is no difference when compared to the previous case: PS volume $V_s$ still has the rectangular shape and $n=1$ (Fig.~\ref{parallelogram}c).

For instruments using the beam focusing on the detector (see e.g.~\cite{kentzinger2004}) the detector pixel can be considered as the “sample”, while the real sample can be placed anywhere between the detector and the NTS exit.

%% file: guide_exit.tex
\subsection{Sample illumination}
\label{samp_illum}

\begin{figure}[h!]
\begin{center}
\begin{minipage}[h]{\linewidth}
\centering
\includegraphics[width=1\linewidth]{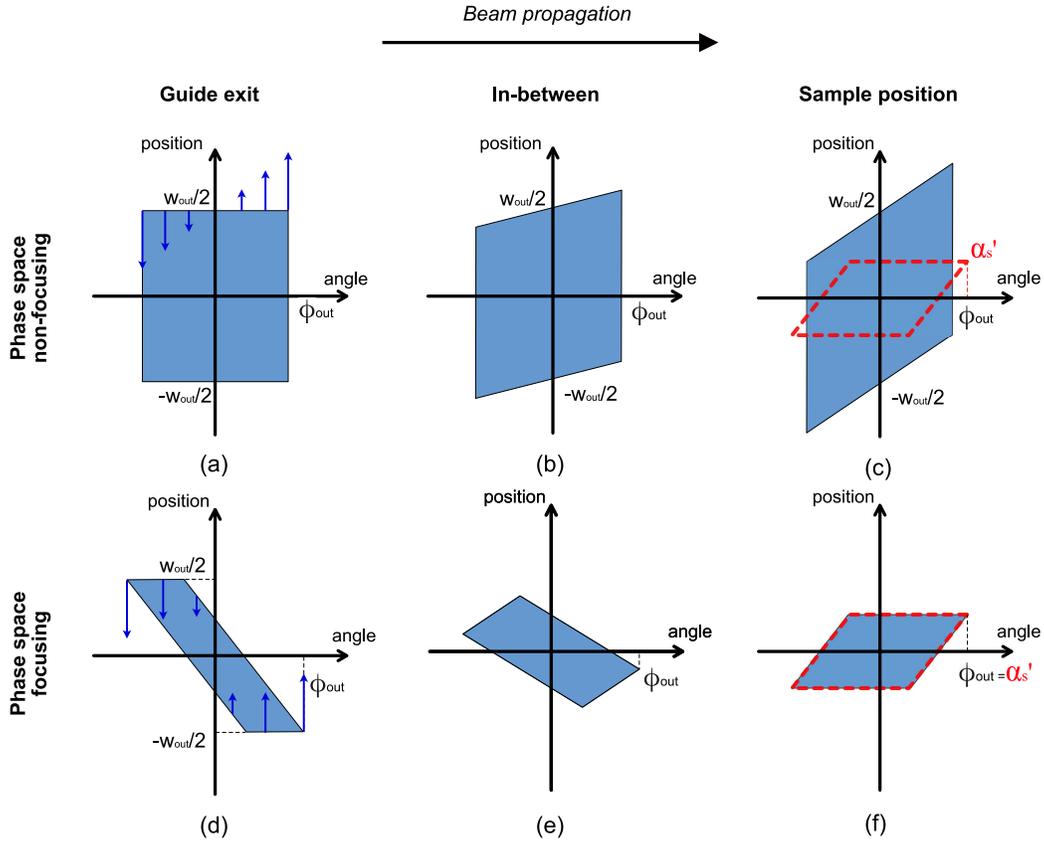}
\end{minipage}

\caption{Evolution of the shape of beam PS volume between NTS exit and sample position for non-focusing case (upper row (a--c)) and focusing case (lower row (d--f)). Dark blue arrows in panels (a) and (d) show the direction where points move when the beam propagates, length of arrows corresponds to the speed of this movement. The PS volume required by the instrument (shown with red line) is similar to one depicted in Fig.~\ref{parallelogram}b.}
\label{phase_space_big}
\end{center}
\end{figure}

Let us consider the shape of PS volume of the neutron beam  $V_{out}$ at the NTS exit (Fig.~\ref{phase_space_big}). Two extreme cases can be distinguished: 
\begin{enumerate}
    \item Phase space non-focusing (PS NF) exit of NTS, when there is no correlation between angle and position of each neutron trajectory (Fig.~\ref{phase_space_big}a);
    \item Phase space focusing (PS F) exit of NTS: PS volume  $V_{out}^{F}$ at the NTS exit has a special shape (position-angle correlation), so that after propagation to the sample position the shape of $V_{out}^F$ matches $V_s$ (Figs.~\ref{phase_space_big}d,f).
\end{enumerate}

Note that we speak of focusing in phase space, not in real one. Focusing here and further on refers to matching of two PS volumes shapes.

Consider case 1 in more details. PS volume at the PS non-focusing exit of the NTS has rectangular shape (Fig.~\ref{phase_space_big}a) and can be calculated as 

\begin{equation}
    V_{out}^{NF} = 2 w_{out} \phi_{out}.
    \label{volume_exit_nf}
\end{equation}

In other words neutron trajectories going through a given point at the NTS exit may have any angle with the NTS axis within the divergence $\phi_{out}$. Figs.~\ref{phase_space_big}a--c show the transformation of PS volume shape between the NTS exit and sample. As neutrons are moving along trajectories with a given angle, their space coordinates are changed proportionally to this angle (see dark blue arrows) and the length of their trajectory, so that the rectangular shaped PS volume is transformed into the parallelogram shaped one. In agreement with Liouville theorem the volume itself is conserved, only its shape is transformed.

The PS volume $V_s$ required by the instrument is shown in Fig.~\ref{phase_space_big}c by red dashed line (see also Fig.~\ref{parallelogram}). Optimal instrument performance is reached if $V_s$ is fully inscribed in the transformed $V_{out}^{NF}$, and the excess volume $V_{out}^{NF} - V_s$ is minimal. In other words, these conditions correspond to the full sample illumination and minimum of background and using ~(\ref{factor}) and (\ref{factor2}) we can write:

\begin{equation}
    \phi_{out} = \alpha_s' = n \alpha_s,
    \label{1st_req}
\end{equation}

\begin{equation}
    w_{out} = 2L_{out}\alpha_s'' + d_s = 2L_{out}(2-n)\alpha_s + d_s.
    \label{full_illumination}
\end{equation}

These two conditions are important for further discussion and we will refer to them as Optimal and Full Sample Illumination (OFSI) conditions. Practically these conditions allow one to minimise the over-illumination of the sample. The over-illumination can lead to following consequences:
\begin{enumerate}
\item Neutrons with too high angles hit the sample, that worsens the resolution and violates basic instrument requirements. This is prevented by condition~(\ref{1st_req}).
\item Some neutrons reach the sample position outside of the sample. They can be scattered at the sample holder that results in unwanted background at the detector and reduce the signal-to-background ratio. The number of such neutrons is minimised by condition~(\ref{full_illumination}), while still providing full sample illumination.
\end{enumerate}

A non-optimal situation, where both OFSI conditions are violated, is shown in Fig.~\ref{phase_space_big}c. The sample is under-illuminated since $\phi_{out} < \alpha_s'$, that provides better than required resolution however with decreased sample flux. Simultaneously $w_{out}$ is larger than required, leading to a high proportion of “useless” neutrons at the sample position.

If both OFSI conditions (\ref{1st_req}) and (\ref{full_illumination}) are met, we can write

\begin{equation}
    V_{out}^{NF} = 2 (2L_{out}(2-n)\alpha_s + d_s) n \alpha_s
    \label{V_NF}
\end{equation}
Thus, the optimal instrument performance can be achieved, only if this equation for PS volume at the NTS exit holds true.

Another extreme case of the NTS exit is the PS focusing one (PS F), which brings to the sample the exact PS volume, required by the instrument, and this volume depends neither on $L_{out}$ nor on $n$:

\begin{equation}
    V_{out}^{F} = V_s = 2d_s\alpha_s.
\label{V_F}
\end{equation}
In this case the OFSI conditions~(\ref{1st_req}) and~(\ref{full_illumination}) are also held, despite they were initially derived for the case of PS non-focusing case. 

Figs.~\ref{phase_space_big}d--f show the evolution of the shape of PS volume provided by such NTS exit and allow to reconstruct what type of position-angle correlation is required at the NTS exit. 

The great diversity of modern NTSs does not allow for an immediate answer which of them have PS focusing or non-focusing exits. As an obvious example of NTS with phase space NF exit one can consider a straight neutron guide. An optimized elliptic focusing guide could be NTS with phase space F exit. One should check the PS volume of the beam at the sample position to determine precisely the NTS exit type. Certainly, the shape of PS volume from a realistic NTS is constrained by two extreme cases described above in Eqs.~(\ref{V_NF},\ref{V_F}).

%% file: guide_entrance.tex
\newpage
\subsection{NTS entrance illumination}

\begin{figure}[t]
\begin{center}
\begin{minipage}[h]{0.49\linewidth}
\centering
\includegraphics[width=1\linewidth]{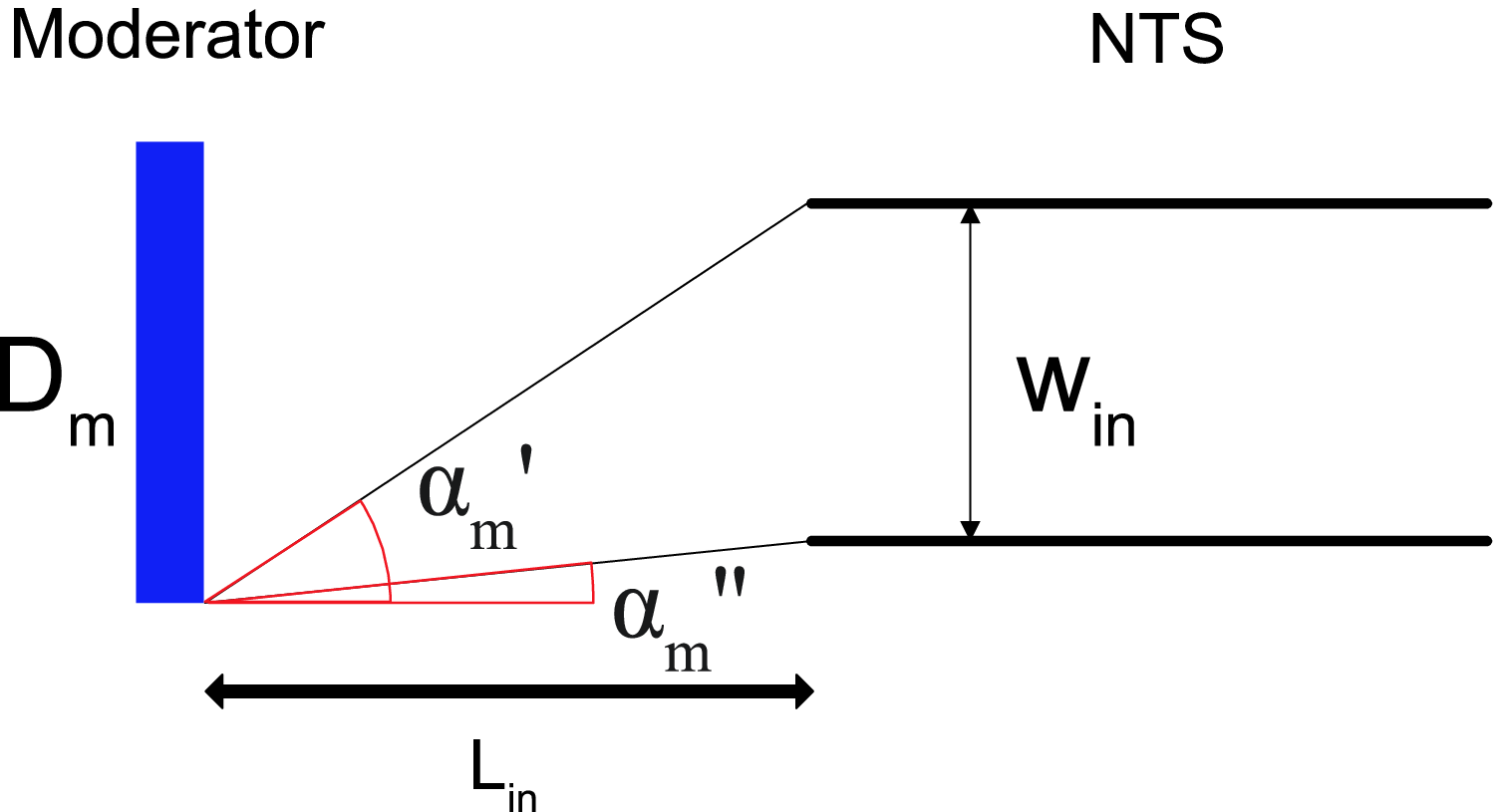} \\ (a)
\end{minipage}
\begin{minipage}[h]{0.49\linewidth}
\centering
\includegraphics[width=1\linewidth]{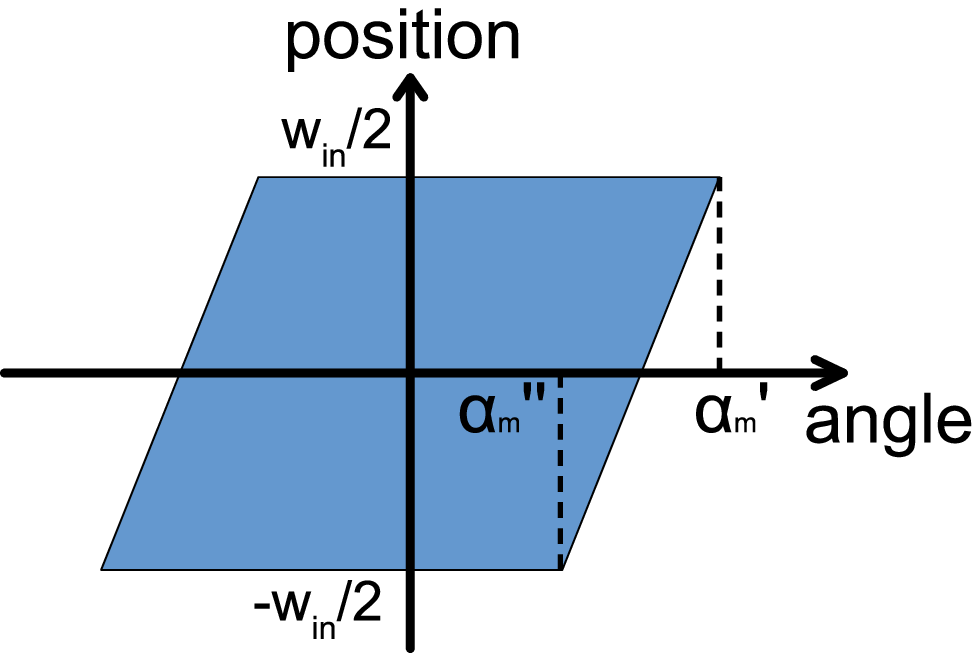} \\ (b)
\end{minipage}
\caption{(a) NTS entrance illumination scheme. (b) PS volume $V_m$ provided by the moderator given the distance $L_{in}$ and NTS entrance size $w_{in}$.}
\label{simple_source}
\end{center}
\end{figure}

NTS entrance illumination scheme is shown in Fig.~\ref{simple_source}a. A homogeneous and isotropic neutron source of size $D_m$  is separated from the NTS entrance of size $w_{in}$ by the distance $L_{in}$. While the moderator emits neutrons in all directions we are only concerned with those hitting the NTS entrance. Geometry shown in Fig.~\ref{simple_source}a defines the collimation of the incident neutron beam. At the distance $L_{in}$ from the moderator surface the PS volume $V_m$, which can be potentially accepted by the NTS entrance, has a shape of parallelogram, as shown in Fig.~\ref{simple_source}b.

Angles $\alpha_m'$ and $\alpha_m''$ are defined as follows:

\begin{equation}
    \alpha_m' = \dfrac{D_m + w_{in}}{2L_{in}},
    \label{div_source1}
\end{equation}

\begin{equation}
    \alpha_m'' = \dfrac{D_m - w_{in}}{2L_{in}}.
    \label{div_dource2}
\end{equation}

Then PS volume $V_m$  can be calculated as

\begin{equation}
    V_m = w_{in} (\alpha_m' + \alpha_m'') = w_{in}\dfrac{D_m}{L_{in}}.
    \label{moderator_volume}
\end{equation}

Depending on e.g. the shape and coating of the neutron guide the NTS can accept this PS volume either fully or partially. As in the previous section, one can consider two extreme cases: with or without PS focusing. The latter corresponds to the NTS entrance, which accepts rectangular PS volume without any position-angle correlations, while the former corresponds to a specific NTS entrance with the acceptance exactly matching PS volume shown in Fig.~\ref{simple_source}b.

\begin{figure}[h]
\begin{center}
\begin{minipage}[h]{\linewidth}
\centering
\includegraphics[width=1\linewidth]{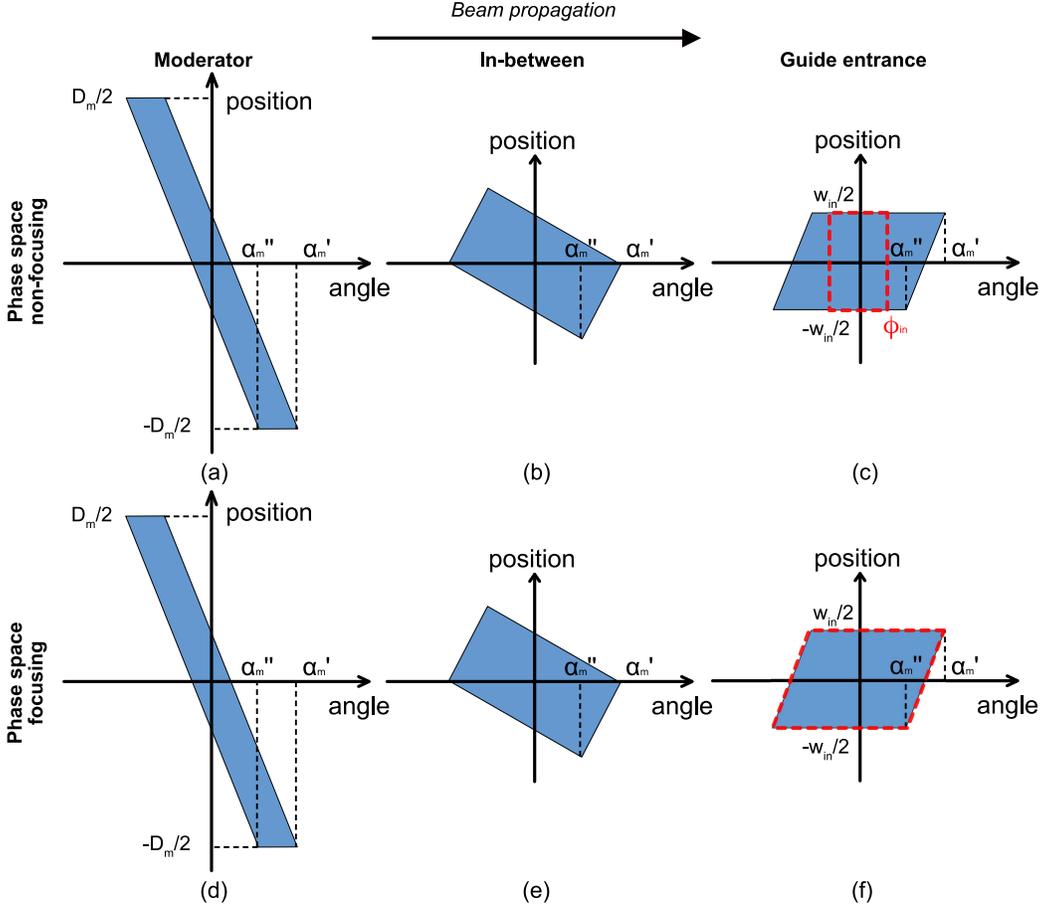}
\end{minipage}

\caption{Evolution of the PS volume $V_m$ shape between the moderator (a,d) and the NTS entrance (c,f). First row a--c shows non-focusing case, second row d--f shows focusing case. Shown with red dashed line is the PS volume accepted by the NTS.}
\label{phase_space_big2}
\end{center}
\end{figure}

The evolution of the PS volume $V_m$ between the moderator and the NTS entrance is shown in Fig.~\ref{phase_space_big2}. The shape of the PS volume delivered to the NTS entrance is the same for both PS NF and PS F cases, however the shape of accepted PS volumes is different as shown by red dashed lines in Fig.~\ref{phase_space_big2}c and Fig.~\ref{phase_space_big2}f, respectively.

PS non-focusing NTS entrance only accepts PS volume (see Fig.~\ref{phase_space_big2}c):

\begin{equation}
    V_{in}^{NF} = 2 w_{in} \phi_{in}.
    \label{volume_nf}
\end{equation}
If there is excessive incident PS volume, which comprises neutron trajectories entering the NTS under angles exceeding the critical angle of the guide walls coating for corresponding wavelengths, it will be absorbed in the coating of the neutron guide walls creating a high-energy gamma-background, so that quite bulky and rather expensive radioactive shielding around the NTS will be required. This background can be minimized by imposing the condition ~(\ref{div_equ}) that still allows for the full illumination of the NTS entrance (see Fig.~\ref{phase_space_big2}c):

\begin{equation}
    \phi_{in} = \alpha_m''.
    \label{div_equ}
\end{equation}

Then from Eq.~(\ref{div_dource2}), we can obtain the optimal moderator size $D_{opt}$:

\begin{equation}
    D_{opt} = w_{in} + 2 L_{in} \phi_{in}.
    \label{mod_size_nf}
\end{equation}
If the actual source size deviates from this value, then either the NTS entrance is over-illuminated leading to an additional background along the transport system, or it is under-illuminated leading to reduced instrument performance.

Now it is possible to rewrite the expression~\eqref{volume_nf} for PS volume:

\begin{equation}
    V_{in}^{NF} = w_{in} \dfrac{D_{opt}-w_{in}}{L_{in}}.
\end{equation}
Note that in the case of PS non-focusing NTS entrance the optimal moderator size $D_{opt}$ is always larger than the NTS entrance size $w_{in}$.

Another extreme case is that of the PS focusing NTS entrance (Fig.~\ref{phase_space_big2}f), for which
\begin{equation}
    V_{in}^{F} = V_m.
\end{equation}
The optimal moderator size then can be calculated from Eq.~(\ref{moderator_volume}). In this case optimal moderator can be of any size relative to the NTS entrance, including smaller than that.

In practice, all known to us modern NTSs have the PS non-focusing entrance, where $\phi_{in}$ is determined by critical momentum transfer of guide walls. At the moment, we don't have any suggestions for the construction of optical system with such position–angle correlation at its finite size entrance as shown in Fig.~\ref{phase_space_big2}f.

%% file: COFSI.tex
\subsection{Moderator size for optimal and full sample illumination}

A NTS with PS focusing properties at the entrance does not necessarily possess them at the exit. One can distinguish four extreme cases that are the combinations of PS focusing/non-focusing properties at the entrance and exit of the NTS. We notate them with two symbols, referring to the entrance and the exit, respectively: NF--F, NF--NF, F--NF and F--F. For example, the NTS of the NF--F type possess PS non-focusing entrance and PS focusing exit.

In case of ideal transport $V_{in} = V_{out}$ (see Eq.~(\ref{ideal_trans})). Then we can tie together instrument parameters and the optimal moderator size. Collecting corresponding expressions from previous subsections we obtain:

\begin{enumerate}
    \item[a)] for NF--NF type NTS\begin{equation}
   w_{in} \dfrac{D_{opt} - w_{in}}{L_{in}}  = V_{in}^{NF} = V_{out}^{NF} = 2n\alpha_s(2L_{out}\alpha_s(2-n) + d_s);
   \label{KOPZO_NF1}
\end{equation}

\item[b)] for F--NF \begin{equation}
     w_{in}\dfrac{D_{opt}}{L_{in}} = V_m= V_{in}^{F} = V_{out}^{NF} = 2n\alpha_s(2L_{out}\alpha_s(2-n) + d_s);
     \label{KOPZO_F1}
\end{equation}

\item[c)] for NF--F \begin{equation}
     w_{in} \dfrac{D_{opt} - w_{in}}{L_{in}}  = V_{in}^{NF} = V_{out}^{F} = V_s = 2 d_s \alpha_s;
     \label{KOPZO_NFF}
\end{equation}

\item[d)]  for F--F \begin{equation}
     w_{in}\dfrac{D_{opt}}{L_{in}} = V_m= V_{in}^{F} = V_{out}^{F} = V_s = 2 d_s \alpha_s.
    \label{KOPZO_F2}
\end{equation}
\end{enumerate}

From here it is possible to derive functions $D_{opt}(w_{in})$, linking optimal moderator size and NTS entrance size via instrument parameters. We call these functions Curves of Optimal and Full Sample Illumination (COFSIs), where “optimal” refers to minimal background and “full” refers to maximal sample flux. Table~\ref{KOPZO_types_simple} contains expressions for COFSIs obtained for different types of NTSs and types of collimation before the sample: for the double-slit collimator (using  Eq.~\eqref{db-coll}) and for Soller or natural collimation ($n=1$).

\begin{table}[h]
\begin{center}
\begin{tabular}{|c|c||c|c|}
\hline
\multicolumn{2}{|c||}{\multirow{2}{*}{$D_{opt} = $}} & \multicolumn{2}{|c|}{
    \textit{NTS entrance}} \\
\cline{3-4}
\multicolumn{2}{|c||}{} & F & NF \\
\hline
\hline

\multirow{6}{*}{\rotatebox{90}{\textit{NTS exit}}} & F & \rule{0pt}{23pt}$\dfrac{2 d_s \alpha_s L_{in}}{w_{in}}$ & $\dfrac{2 d_s \alpha_s L_{in}}{w_{in}} + w_{in}$ \\

& \begin{tabular}{@{}c@{}}NF \\ \footnotesize{double-slit}\end{tabular} & $\dfrac{2 d_s \alpha_s L_{in}}{w_{in}} \cdot \dfrac{n}{n-1}$ & \rule{0pt}{30pt} $\dfrac{2 d_s \alpha_s L_{in}}{w_{in}}\cdot \dfrac{n}{n-1} + w_{in}$ \\ [15pt]

& \begin{tabular}{@{}c@{}}NF \\ \footnotesize{Soller or natural}\end{tabular} & $\dfrac{2 d_s \alpha_s L_{in}}{w_{in}} + \dfrac{4\alpha_s^2L_{in}L_{out}}{w_{in}}$ & $\dfrac{2 d_s \alpha_s L_{in}}{w_{in}} + \dfrac{4\alpha_s^2L_{in}L_{out}}{w_{in}} + w_{in}$  \\ [15pt]
\hline
\end{tabular}
\caption{ COFSIs for four extreme cases of NTS.}
\label{KOPZO_types_simple}
\end{center}
\end{table}

\begin{figure}[h!]
\begin{center}
\begin{minipage}[h]{0.7\linewidth}
\centering
\includegraphics[width=1\linewidth]{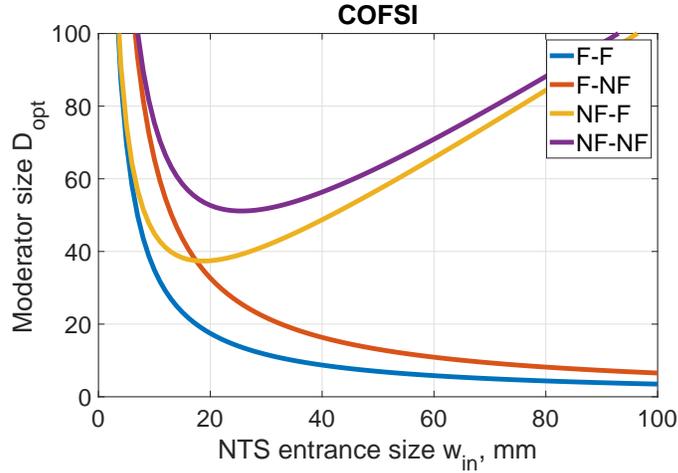}
\end{minipage}
\caption{Example of COFSIs for the particular neutron instrument (see parameters in the text) for four extreme cases of NTS.}
\label{COFSI_example}
\end{center}
\end{figure}

As an example, four extreme cases of COFSIs for the instrument with following parameters: $d_s=10$~mm, $\alpha_s = 0.5$\textdegree{}, $n=1$, $L_{in} = 2000$~mm and $L_{out}=500$~mm are shown in Fig.~\ref{COFSI_example}. 

For any particular type of NTS one can calculate COFSIs basing only upon instrument parameters and nothing else. COFSIs allow to find optimal moderator size for any NTS entrance size. Such moderator provides three important advantages for the instrument: 
\begin{enumerate}
    \item full sample illumination (in all cases), allowing to achieve maximal sample flux for given instrument parameters and moderator brilliance;
    \item minimal (for PS NF exit) or none (for PS F exit) background at the sample position, allowing to improve signal-to-noise ratio and detect weak signals;
    \item minimal (for PS NF entrance) or none (for PS F entrance) background along the NTS, thus reducing the shielding cost.
\end{enumerate} 
For each of NTS types there is only one optimal moderator size for given NTS entrance size. Any deviation from the optimal moderator size leads to the loss of above mentioned advantages (this is considered in details in Sec.~\ref{sample_flux}).

As shown in Fig.~\ref{COFSI_example} for the NTSs with PS focusing entrance COFSIs are hyperbolas (red and blue curves) and for the NTSs with PS non-focusing entrance they are the sum of hyperbolic and linear functions (yellow and violet curves). One can see that PS F entrance allows for the optimal use of very small moderators (in the range of tenths of mm) with very high brilliance, such as narrow para-H$_2$ moderators.

Consider NTSs with PS non-focusing entrance. Corresponding COFSIs define the minimal optimal moderator size $D_{opt}^{min}$, which is obtained for NTS entrance size $w_{in}^{min}$. For any smaller moderator it is impossible to achieve full and optimal sample illumination. It will be shown in Sec.~\ref{para-h2} that this size should be as small as possible for an efficient use of para-H$_2$ moderators.

NTS with PS non-focusing entrance may have either PS focusing or non-focusing exit. In case of NF--F NTS the minimum is reached when (see Table~\ref{KOPZO_types_simple})

\begin{equation}
    w_{in}^{min} = \sqrt{2d_s\alpha_sL_{in}}.
    \label{w_min1}
\end{equation}
In case of NF--NF NTS with double-slit collimation
\begin{equation}
    w_{in}^{min} = \sqrt{2d_s\alpha_sL_{in}\dfrac{n}{n-1}}
    \label{w_min_double_slit}
\end{equation}
and for NF--NF NTS with Soller or natural collimation
\begin{equation}
    w_{in}^{min} = \sqrt{2d_s\alpha_sL_{in} + 4\alpha_s^2L_{in}L_{out}}.
    \label{w_min}
\end{equation}

Substituting $w_{in}^{min}$ from Eqs.~(\ref{w_min1}--\ref{w_min}) into corresponding expressions in Table~\ref{KOPZO_types_simple} we obtain that in all cases:
\begin{equation}
    D_{opt}^{min} = 2 w_{in}^{min}.
    \label{D_min}
\end{equation}

\begin{figure}[t]
\begin{center}
\begin{minipage}[h]{0.49\linewidth}
\centering
\includegraphics[width=1\linewidth]{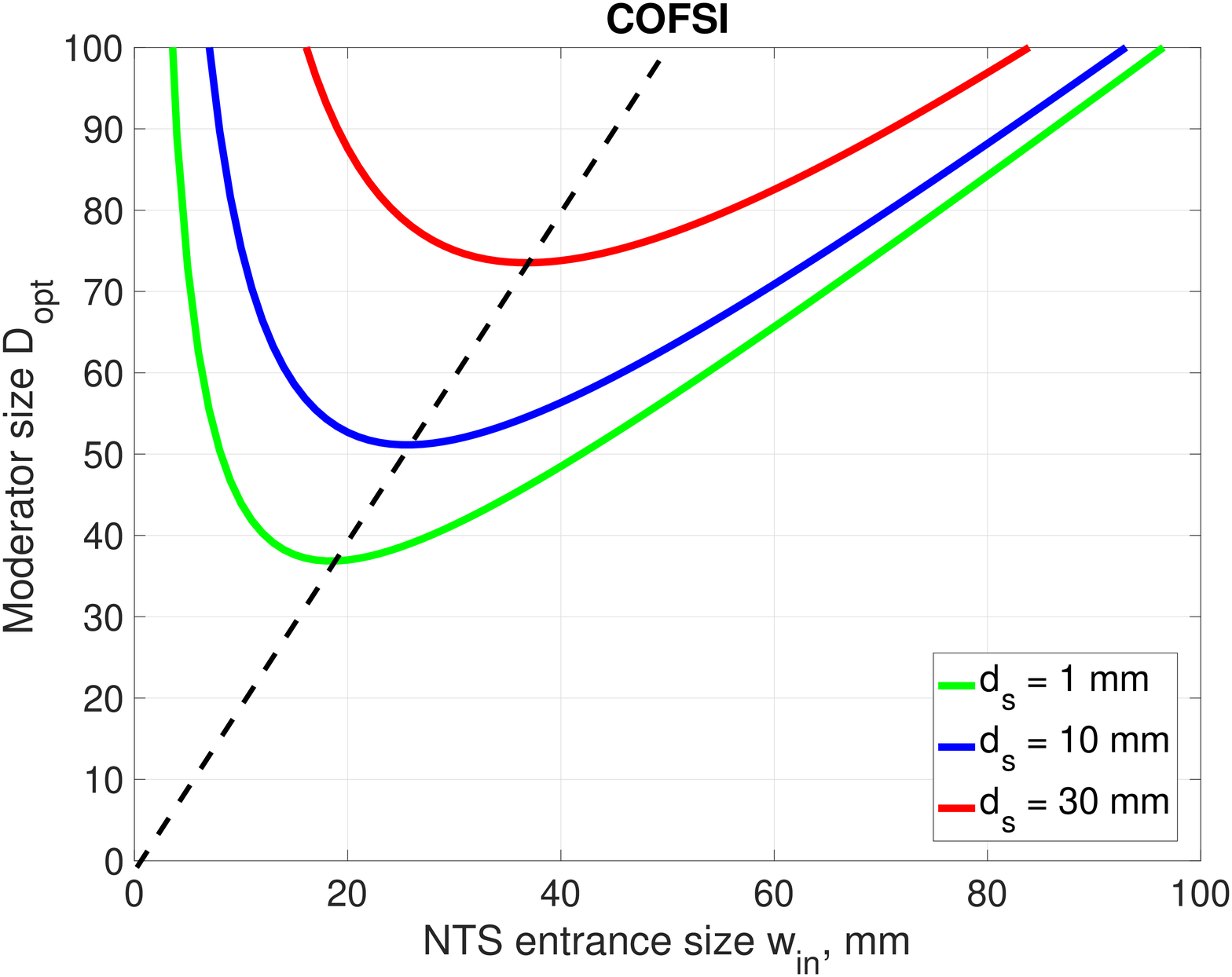} \\ (a)
\end{minipage}
\begin{minipage}[h]{0.49\linewidth}
\centering
\includegraphics[width=1\linewidth]{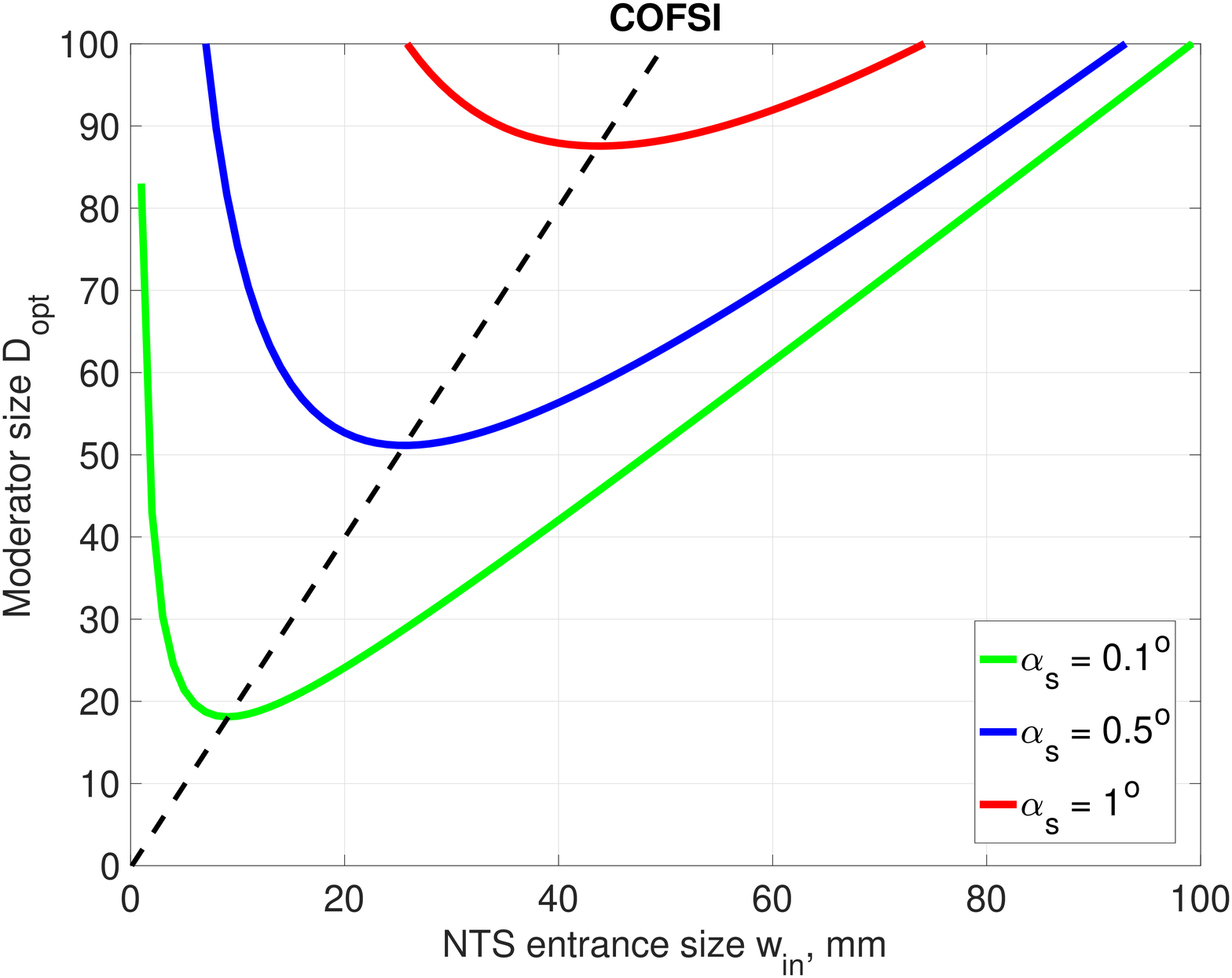} \\ (b)
\end{minipage}
\caption{COFSIs for NF-NF NTS of the instrument with $L_{in} = 2000$~mm, $L_{out}=500$~mm: (a) with $\alpha_s=0.5$\textdegree{} and varied $d_s$; (b) with $d_s=10$~mm and varied $\alpha_s$. Dashed line corresponds to $D_{opt} = 2w_{in}$, on which all COFSI minima are located.}
\label{COFSI_variable}
\end{center}
\end{figure}

COFSIs for different values of sample size $d_s$ and required resolution $\alpha_s$ are presented in Fig.~\ref{COFSI_variable}. All COFSIs minima are on one line corresponding to $D_{opt} = 2w_{in}$ (shown as dashed line). Neutron instruments with smaller $d_s$ or $\alpha_s$ perform optimally with smaller moderators.

Distance $L_{in}$ between moderator and NTS entrance also can be minimised to reduce minimal optimal moderator size. At high-flux sources $L_{in}$ is about 1.5--2 m because of the potential radiation or thermal damage to neutron optics. Much smaller  $L_{in}$ are accessible at compact neutron sources.

%% file: sample_flux_analytic.tex
\section{Deviations from the optimal size of moderator}
\label{sample_flux}

\subsection{Analytic calculations of sample flux}
\label{sample_flux_anal}

Let us now discuss the case when the point corresponding to any given combination of moderator and NTS entrance sizes is above or below COFSI. Optimal and full sample illumination conditions (\ref{1st_req}) and (\ref{full_illumination}) applied at the sample position are still hold, but the requirement of the optimal moderator size (see Table~\ref{KOPZO_types_simple}) is violated.

\begin{figure}[h]
\begin{center}
\begin{minipage}[h]{0.95\linewidth}
\centering
\includegraphics[width=1\linewidth]{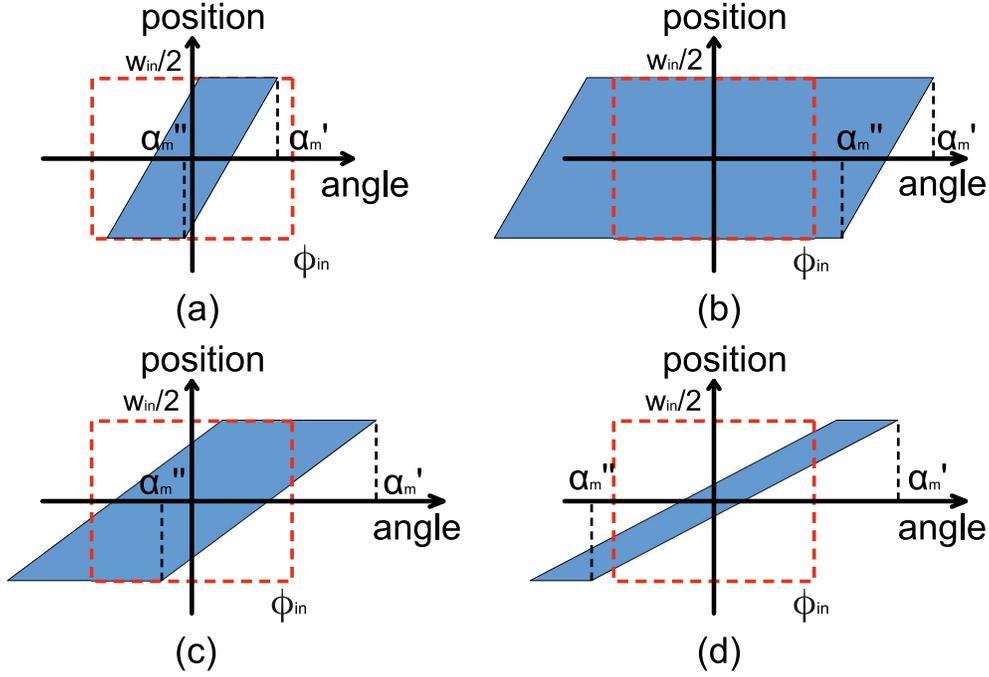}
\end{minipage}

\caption{Comparison of phase volume $V_m$ provided by the moderator (shown in blue) and phase volume $V_{in}$ potentially acceptable by the NTS entrance (inside red dashed line). Only the intersection of those two is accepted by the NTS.}
\label{small_d_source_cap}
\end{center}
\end{figure}

We start with consideration of the NTS entrance illumination (see Fig.~\ref{small_d_source_cap}). For the PS non-focusing entrance the truly accepted PS volume $V_{in}\cap V_m$ is the intersection of the PS volume $V_m$ (shown in blue) illuminating the NTS entrance and PS volume $V_{in}$ (inside red dashed line) potentially acceptable by NTS.

NTS entrance over- or under-illumination is determined solely by how large moderator size $D_m$ is in comparison to NTS entrance size $w_{in}$, that in turn defines divergences $\alpha_m'$ and $\alpha_m''$ (see Fig.~\ref{simple_source}a). Four cases can be distinguished:

\begin{enumerate}
    \item[a)] $\alpha_m'' < \alpha_m' < \phi_{in}$ (Fig.~\ref{small_d_source_cap}a). This case corresponds to the moderator being significantly smaller than optimal. NTS accepts the whole beam (blue area), however is still under-illuminated because the incoming beam divergence is too low:

    \begin{equation}
        V_{in}\cap V_m = V_m = w_{in} (\alpha_m' + \alpha_m'').
    \end{equation}
    
    \item[b)] $\alpha_m' \ge \phi_{in}$ and $\alpha_m'' > \phi_{in}$ (Fig.~\ref{small_d_source_cap}b). This case corresponds to optimal (or larger than optimal) moderator providing high enough beam divergence. NTS accepts the whole potentially acceptable beam (red dashed line) and is fully illuminated (or even over-illuminated):

    \begin{equation}
        V_{in}\cap V_m = V_{in} = 2 w_{in} \phi_{in}.
    \end{equation}

    \item[c)] $\alpha_m' \ge \phi_{in}$ and $-\phi_{in} \le \alpha_m'' \le \phi_{in}$ (Fig.~\ref{small_d_source_cap}c). This is the intermediate case that corresponds to the moderator being slightly less than optimal. NTS accepts only a part of the incident beam because of its too high divergence, however at the same time remains under-illuminated. Given that the parallelogram inclination is equal to $\dfrac{w_{in}}{\alpha_m' - \alpha_m''}$, PS volume accepted by the NTS can be defined as

    \begin{equation}
        V_{in}\cap V_m = 2 w_{in} \phi_{in} - \dfrac{(\phi_{in} - \alpha_m'')^2 w_{in}}{(\alpha_m' - \alpha_m'')}.
    \end{equation}
    
    \item[d)] $\alpha_m' > \phi_{in}$ and $\alpha_m'' < -\phi_{in}$ (Fig.~\ref{small_d_source_cap}d). This is the case of even smaller moderator than in previous case:

    \begin{equation}
        V_{in}\cap V_m = 2 \phi_{in} (\alpha_m' + \alpha_m'') \dfrac{w_{in}}{\alpha_m' - \alpha_m''}.
    \end{equation}
\end{enumerate}

Similarly, we investigate the case of the PS focusing NTS entrance. Note that if the moderator size is optimal $V_m = V_{in}$. If the NTS entrance is under-illuminated (moderator is smaller than optimal), no parts of $V_m$ can be outside of $V_{in}$ and if the NTS entrance is over-illuminated (moderator is larger than optimal), there are no parts of $V_{in}$ not filled with $V_m$. This considerably simplifies the answer for the accepted PS volume:

\begin{equation}
V_{in}\cap V_m =
    \begin{cases}
 V_m, & D < D_{opt} \\
 V_{m,opt}, & D \ge D_{opt},
\end{cases}
\end{equation}
where $V_{m,opt}$ is PS volume provided bu the moderator of optimal size.

If the sample is fully illuminated then according to Eq.~(\ref{brilliance}) sample flux $\Phi_s$ is the product of the PS volume $V_s$ required by the instrument and the brilliance $b_{out}$ of the delivered neutron beam:

\begin{equation}
    \Phi_s = b_{out}V_s.
    \label{sample_flux_eq}
\end{equation}

Note that here the brilliance $b_{out}$ is averaged over the PS volume $V_s$; this is important if the PS volume is not filled uniformly. In our model of the ideal lossless NTS the reason for that can only be under-illumination of its entrance. Any lacuna in potentially accepted PS volume at the NTS entrance is reproduced in some form at its exit. Inhomogeneities in $V_{out}$ are in turn reproduced at the sample position. Only in rare cases these inhomogeneities are not presented in $V_s$, e.g. for the straight neutron guide. Disregarding such cases, one can write:

\begin{equation}
    \dfrac{b_{out}}{B} = \dfrac{V_{in}\cap V_m}{V_{in}},
    \label{inhomo}
\end{equation}
where $B$ is the moderator brilliance.

\begin{figure}[h!]
\begin{center}
\begin{minipage}[h]{0.49\linewidth}
\centering
\includegraphics[width=1\linewidth]{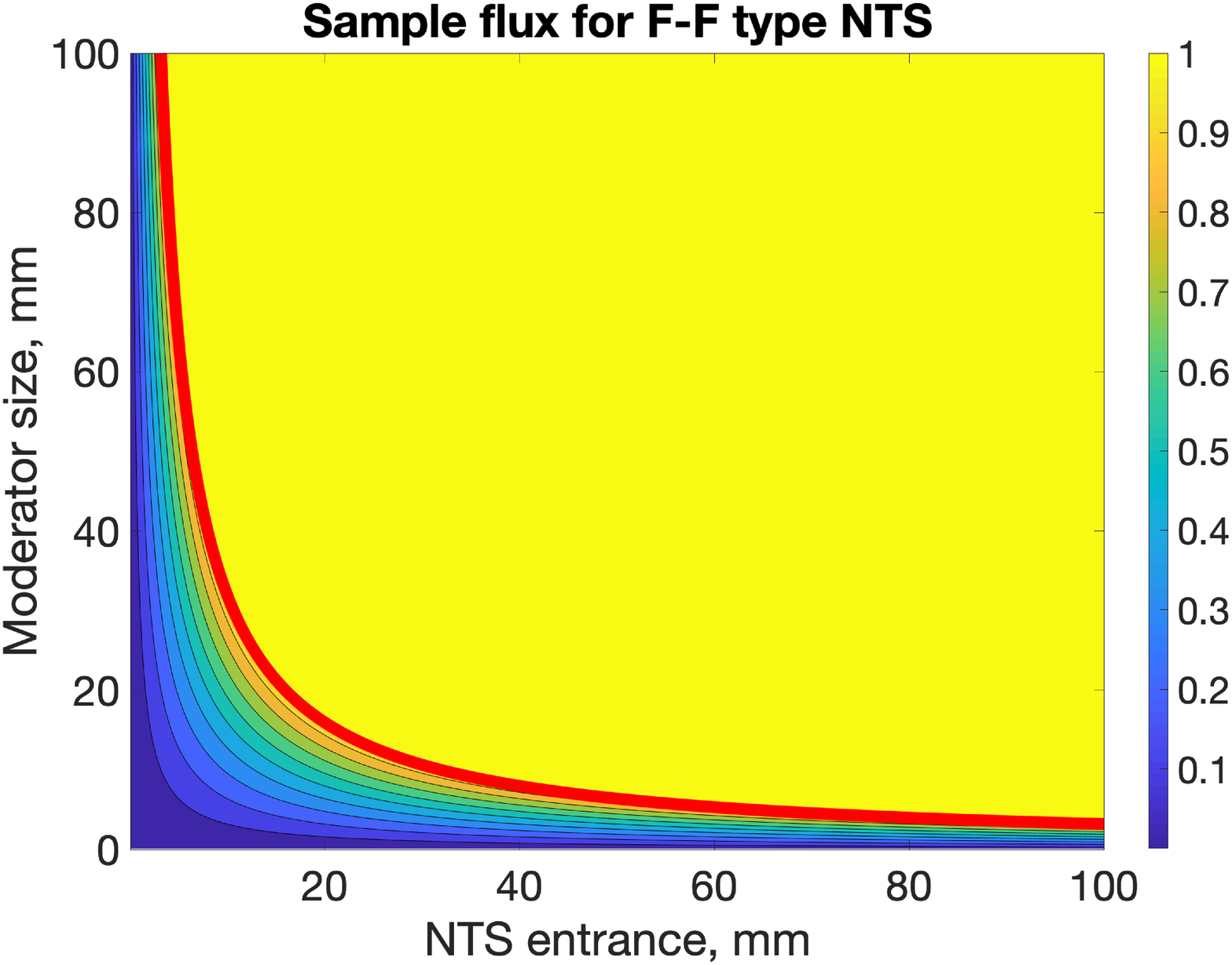} \\ (a)
\end{minipage}
\begin{minipage}[h]{0.49\linewidth}
\centering
\includegraphics[width=1\linewidth]{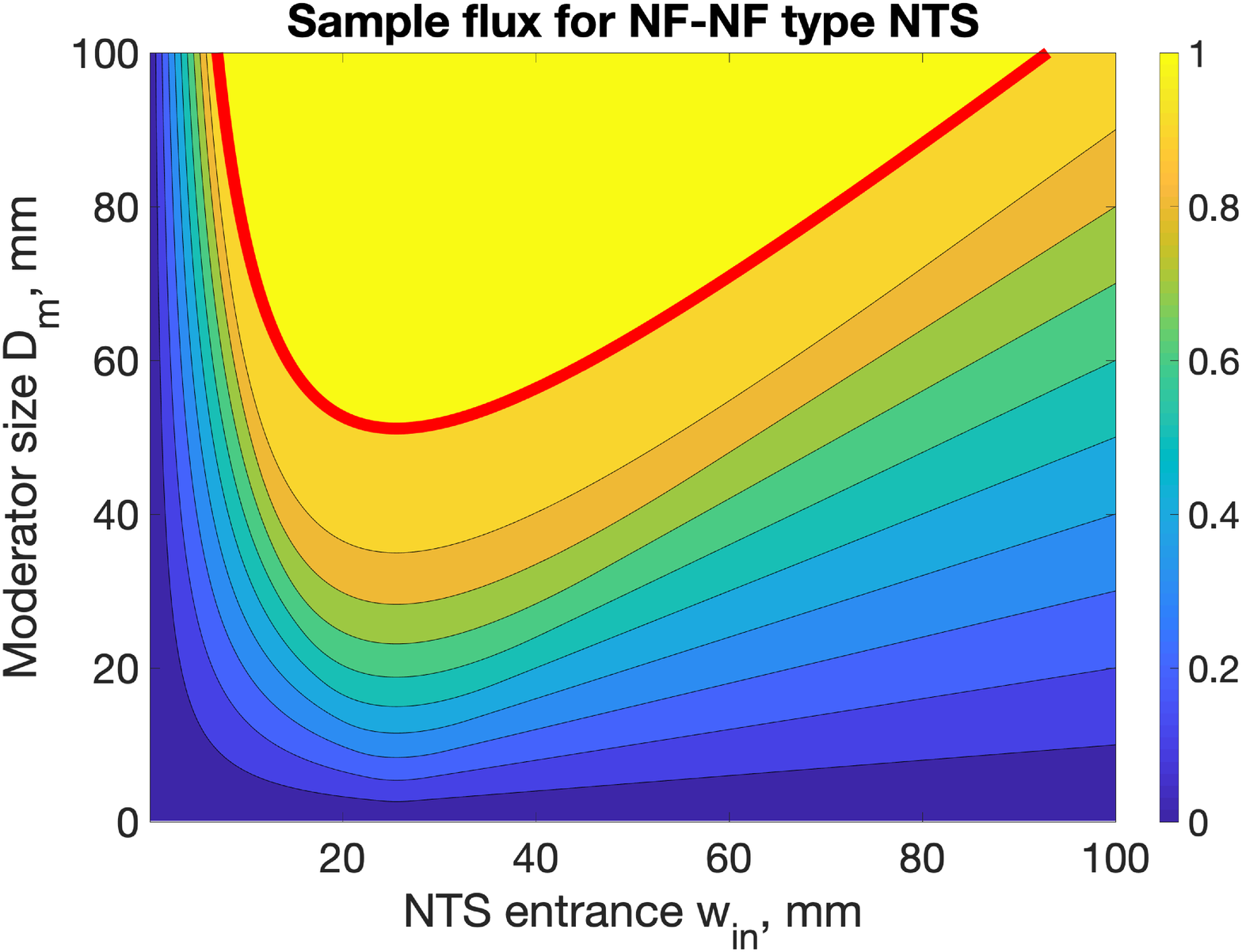}\\ (b)
\end{minipage} 

\caption{Colour maps of sample flux: (a) for NTS with PS focusing entrance; (b) for NTS with PS non-focusing entrance. COFSI is shown in red thick line. Sample flux is normalized on its maximal value independently for both panels.}
\label{Sample_bril_map}
\end{center}
\end{figure}

It is possible now to create colour maps depicting sample flux $\Phi_s$ (Fig.~\ref{Sample_bril_map}). Here instrument parameters are $d_s=10$~mm, $\alpha_s=0.5$~\textdegree, $L_{in} = 2000$~mm and $L_{out}=500$~mm. All points lying on the COFSI (red line) are equivalent in the sense that they provide maximal possible sample flux and minimal background both at the sample position and along the NTS. For points above the COFSI sample flux is still maximal, however there is an increased background along the NTS, that may require an additional shielding. Note that since OFSI conditions are true no additional “useless” are brought to the sample position. For points below the COFSI sample flux is reduced due to the under-illumination of NTS entrance. The COFSI is actually the curve enveloping the region of maximal flux.

%% file: sample_flux_MC.tex
\subsection{Monte-Carlo simulations of sample flux}
\label{sample_flux_mc}

As we defined the NTS as all optical elements positioned between moderator and sample, some of them, like non-ideal neutron guides, crystal monochromators, etc., may introduce transmission losses. Such realistic NTS with brilliance transfer less than unity reduces brilliance at the sample $b_{out}$ compared to the moderator brilliance $B$.

Furthermore, in practice it is not known \textit{a priori} what type of PS focusing possesses particular NTS, for example F--F or NF--F. COFSI for such NTS is somewhere between COFSIs for extreme cases (see Fig.~\ref{COFSI_example}).

To find COFSI for a realistic NTS and corresponding sample flux map it is required to perform Monte-Carlo simulations that allow us to determine the neutron beam brilliance at the sample position for different combinations $(w_{in}, D_m)$. This can be done in two steps:

\begin{enumerate}
    \item For each NTS entrance size $w_{in}$ we select the geometric parameters of the NTS (e.g. elliptic guide foci positions or the mosaicity of crystal monochromator) in such way that the divergence of the beam leaving the NTS is equal to required instrument resolution (condition~(\ref{1st_req})) and beam size at the sample position is just enough to illuminate the sample (condition~(\ref{full_illumination})).
    \item For each $w_{in}$ (i.e. the NTS geometry) the series of Monte-Carlo calculations is performed to obtain sample flux while varying moderator size $D_m$.
\end{enumerate}

We have implemented this algorithm using VITESS simulation package~\cite{zendler2014} and calculated sample flux maps for the instrument with following parameters: $L_{in} = 2000$~mm, $L_{out}=500$~mm, $d_s=10$~mm and $\alpha_s=1$\textdegree. 100 m long neutron guide has elliptic shape and constant $m=3$ wall coating. Calculations are performed for different neutron wavelengths $\lambda$, which define critical angles of reflection from the guide walls. Since the divergence at the exit of the guide is fixed to provide optimal sample illumination, the guide wall inclination depends on $\lambda$. In other words the guide geometry is different not only for each guide entrance size $w_{in}$, but also for each neutron wavelength $\lambda$. 

\begin{figure}[t!]
\begin{center}
\begin{minipage}[h]{0.49\linewidth}
\centering
\includegraphics[width=1\linewidth]{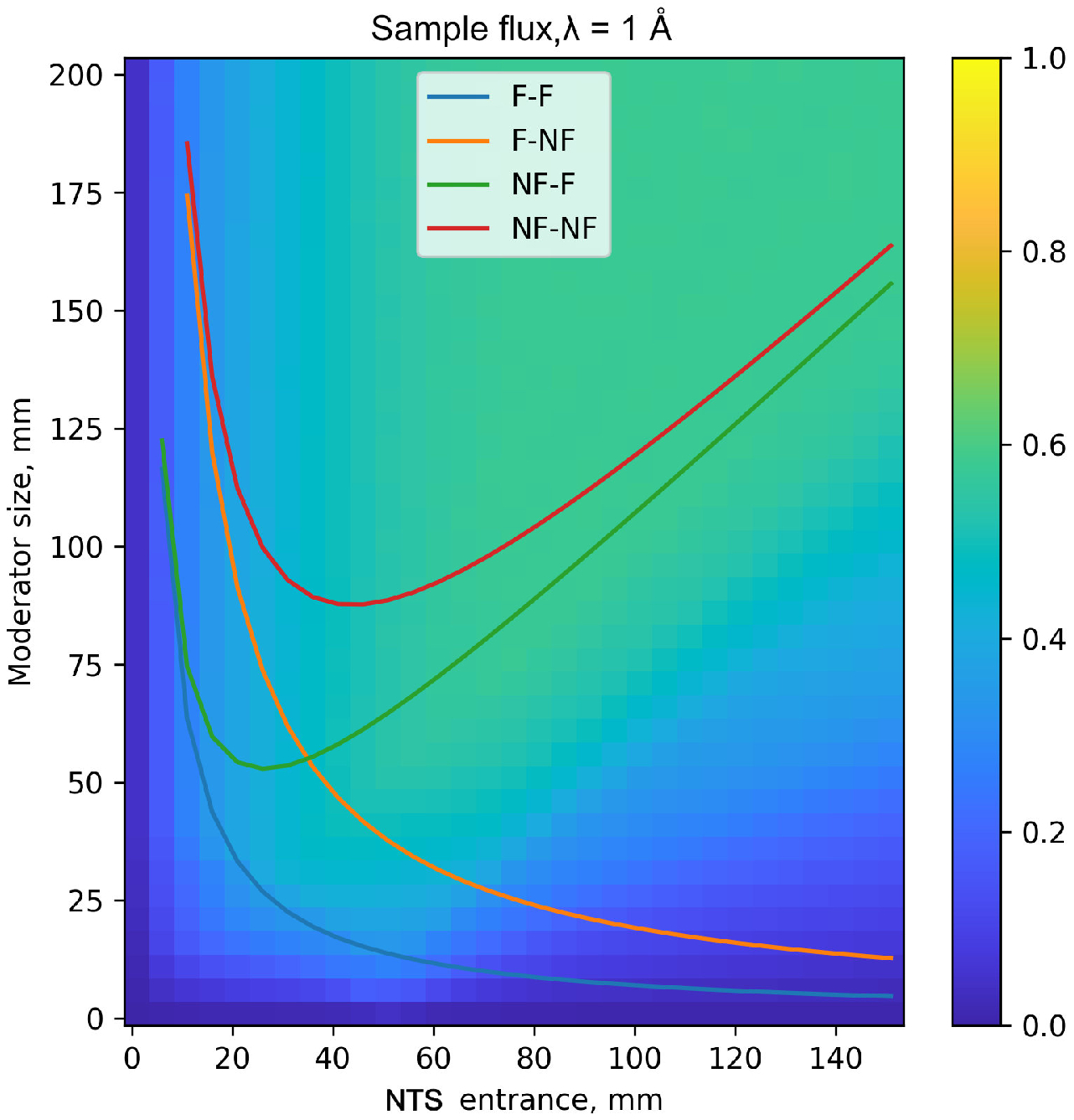} \\ (a)
\end{minipage}
\begin{minipage}[h]{0.49\linewidth}
\centering
\includegraphics[width=1\linewidth]{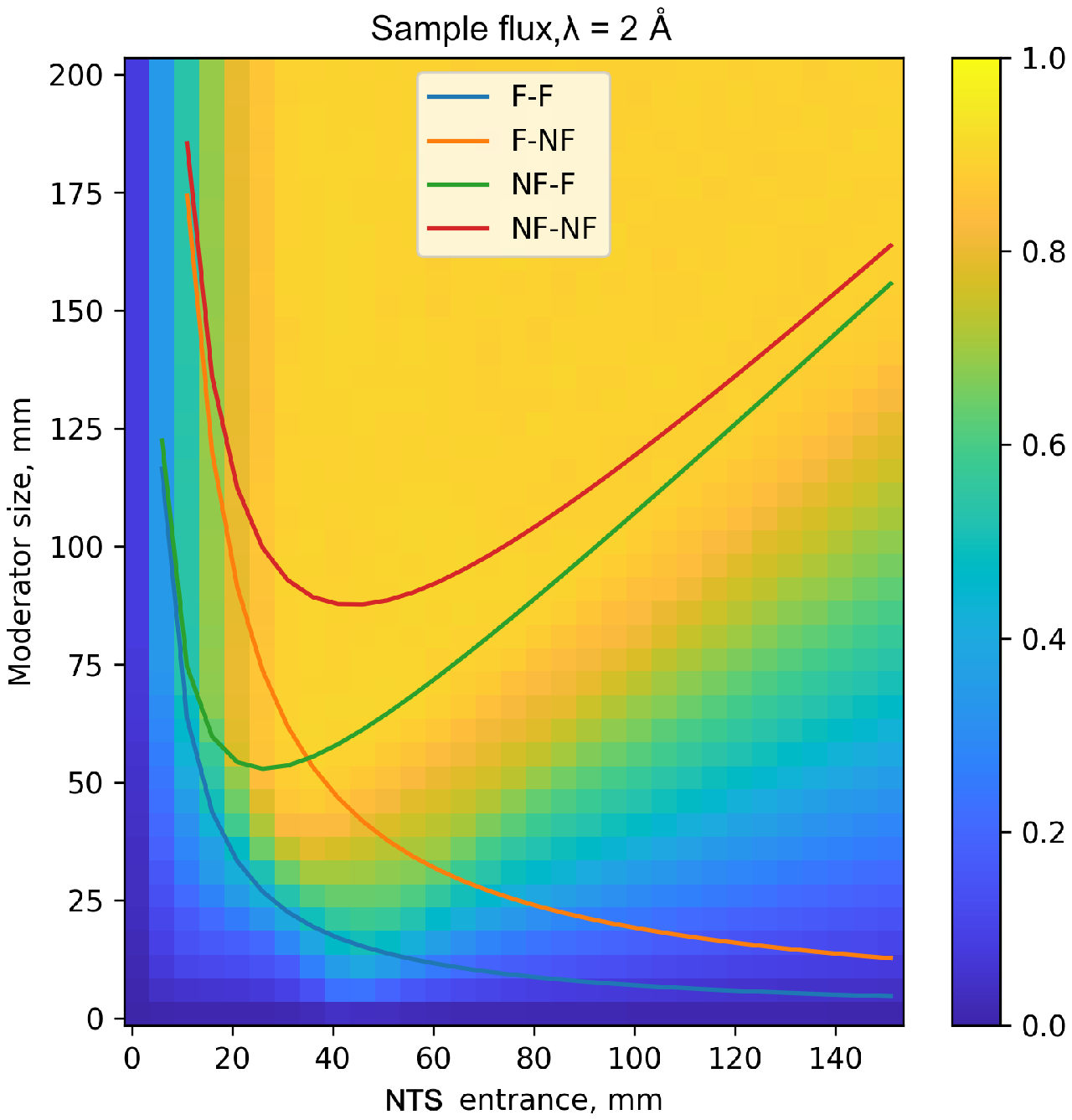}\\ (b)
\end{minipage} 

\begin{minipage}[h]{0.49\linewidth}
\centering
\includegraphics[width=1\linewidth]{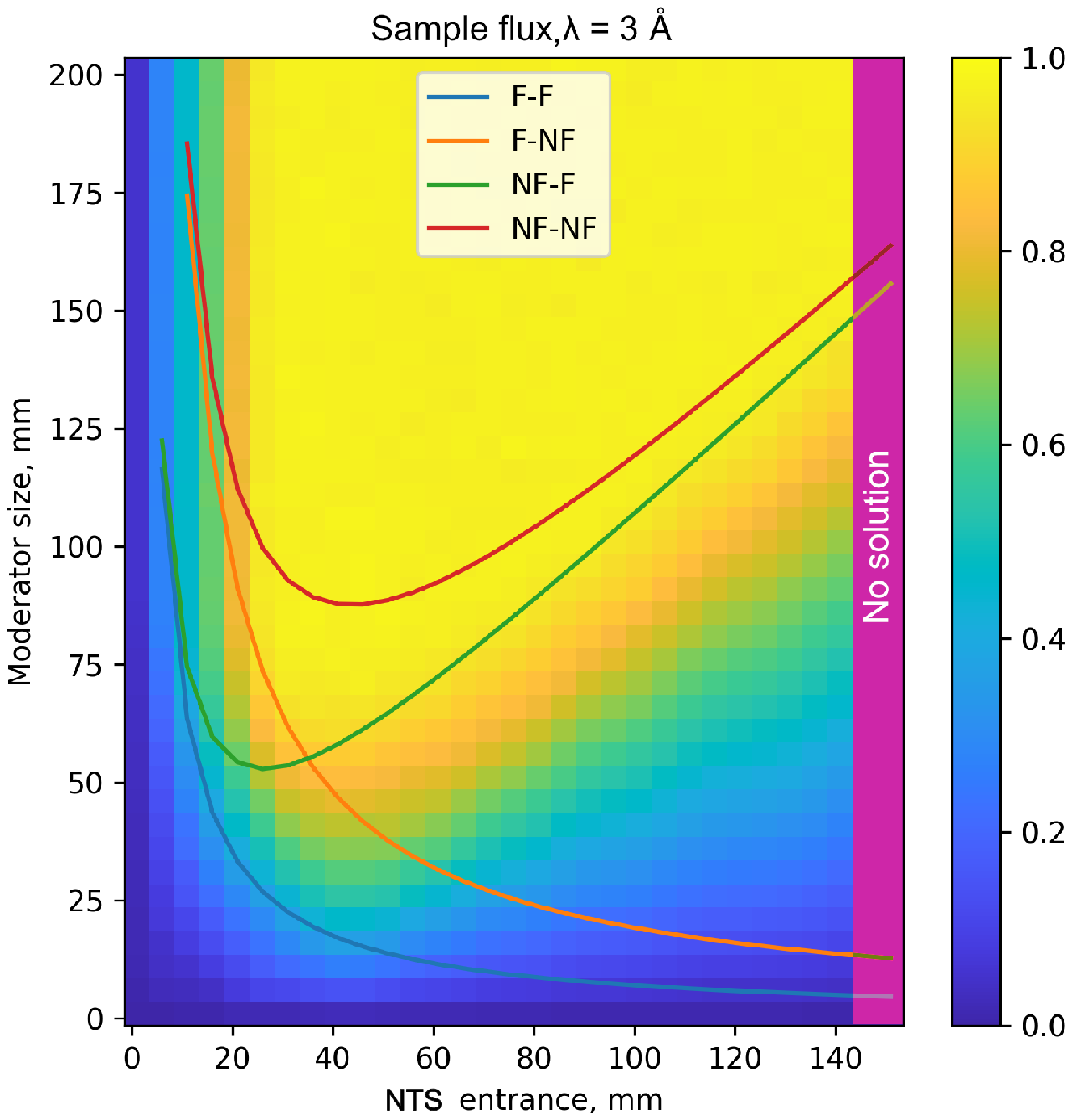} \\ (c)
\end{minipage}
\begin{minipage}[h]{0.49\linewidth}
\centering
\includegraphics[width=1\linewidth]{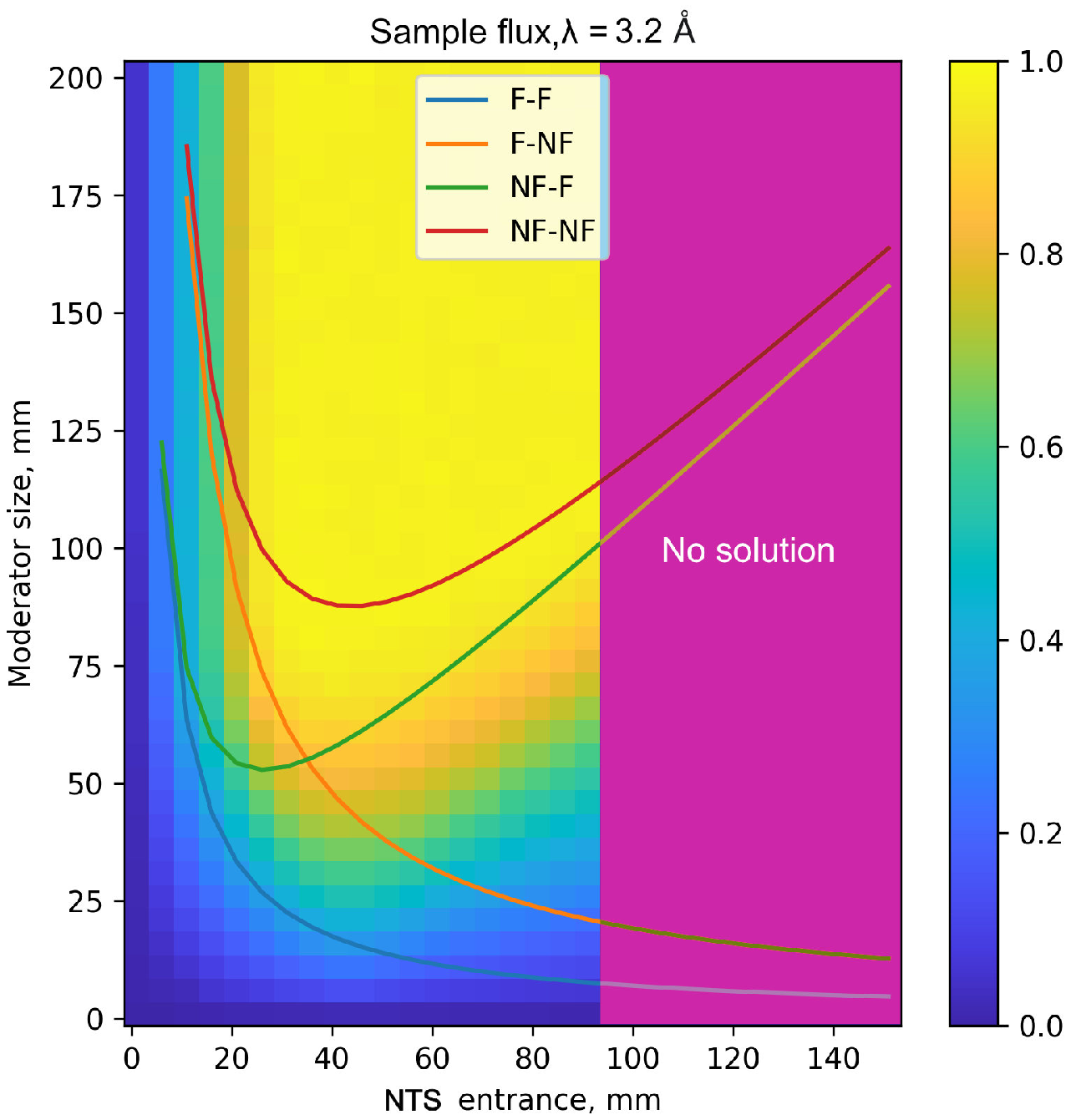}\\ (d)
\end{minipage} 

\caption{Sample flux maps (resolution 5 mm, relative error $\le5\%$) obtained using MC calculations. Sample flux is normalized on the calculated value when $b_{out}=B$. Colour lines depict COFSIs for extreme cases of NTS.}
\label{VITESS_maps}
\end{center}
\end{figure}

Consider simulated sample flux maps shown in Fig.~\ref{VITESS_maps}. They are in a good agreement with our analytical calculations presented earlier in Fig.~\ref{Sample_bril_map}b. As mentioned above in Sec.~\ref{sample_flux_anal} the COFSI envelops the region of maximal flux and such an envelope can now be compared to analytically calculated COFSIs for four extreme cases of neutron optics (shown in solid lines in Fig.~\ref{VITESS_maps}). We can conclude that elliptic guides considered for the simulation are very close to the NTSs of NF--F type.

Thus, the Monte-Carlo analysis allows rather easy classification of realistic NTSs with respect to their phase space focusing properties.

In the region of large $w_{in}$ the inclination of the guide exit wall can be so large that it is impossible to satisfy OFSI conditions~(\ref{1st_req}) and (\ref{full_illumination}). Since beam divergence at the guide exit depends both on the wall inclination and wavelength, this problem should be more pronounced for larger wavelengths. Indeed, for $\lambda=$ 1--2~\AA{} OFSI-compliant guides exist and provide some sample flux for each pair ($w_{in}$, $D_m$) for all $w_{in} \le 150$~mm as shown in Figs.~\ref{VITESS_maps}a,b. 

For increased wavelength of $\lambda=3$~\AA{} the region where OFSI-compliant guides exist  shrinks (limited by the purple stripe in Fig.~\ref{VITESS_maps}c). With increase of wavelength up to $\lambda=3.2$~\AA{} this region shrinks further (Fig.~\ref{VITESS_maps}d), and for $\lambda\ge3.5$~\AA{} disappears, i.e. there are no OFSI-compliant guide solutions at all.

Another deviation from analytical predictions lies in the region of small $w_{in}$. Here the guide must capture relatively large beam divergence and that is the reason why all analytically calculated COFSIs rise up fast here. However, accepted beam divergence for short wavelengths can be too small due to limited critical angle. Comparing sample flux maps for $\lambda=3$~\AA{} and $\lambda=1$~\AA{} (Figs.~\ref{VITESS_maps}c and a), one can see that for large wavelength sample flux quickly rises with increase of the guide entrance size and reaches maximal value $\Phi_s^{max} = 1$, while for shorter wavelength sample flux rises much slower and its maximal value is about $\Phi_s^{max} = 0.6$. Fig.~\ref{VITESS_maps}b  depicts the intermediate case for $\lambda = 2$~\AA{}.

These Monte-Carlo simulations highlight the fact that COFSIs solutions are chromatic. Though the expressions in Table~\ref{KOPZO_types_simple} are derived in a purely geometric way, in practice $\phi_{out}$ usually depends on neutron wavelength. It means that both optimal moderator size and optimal guide geometry are actually different for different wavelengths.

To sum up, Monte-Carlo simulations allow for the generalization of the developed COFSI method by taking into account realistic reflectivity losses (non-ideal neutron transport) and practical guide geometries with various PS focusing properties. Well-designed NTS with minimal transmission losses and high enough reflection angle at its entrance, provides sample flux map very close to analytical predictions from Sec.~\ref{sample_flux_anal}.

%% file: para_hydrogen.tex
\newpage
\subsection{Sample flux in case of low-dimensional para-hydrogen moderator}
\label{para-h2}

Sample flux maps shown in Figs.~\ref{Sample_bril_map},\ref{VITESS_maps} are obtained under the assumption of size-independent moderator brilliance $B$. In practice this corresponds well to large deuterium moderators at research reactors. 

\begin{figure}[t]

\begin{center}
\begin{minipage}[h]{0.5\linewidth}
\centering
\includegraphics[width=1\linewidth]{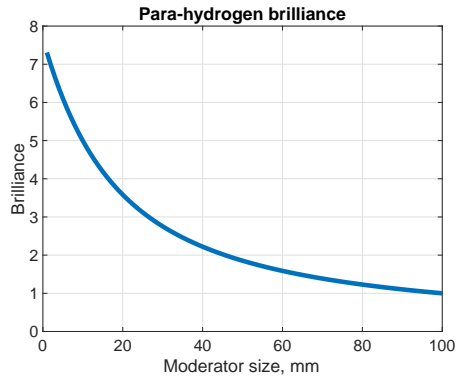}
\caption{Dependence of the brilliance for para-H$_2$ moderator on its size.}
\label{ph2}
\end{minipage}
\end{center}
\end{figure}

However, rather different situation occurs for para-H$_2$ moderators, which exhibit the strong size dependence of brilliance: smaller moderators provide significantly higher brilliance than larger ones~\cite{batkov2013}. For our further considerations we use data from \cite{zanini2019}, Fig. 18, normalized for 100 mm large moderator. We fitted the brilliance with the function

\begin{equation}
    B_{fit}(D_m) = \dfrac{p_1 D_m + p_2}{D_m^2 + q_1D_m + q_2},
    \label{bril_func}
\end{equation}
where $D_m$ is measured in cm, $p_1 = 10.4271$, $p_2=38.3069$, $q_1=3.7588$ and $q_2=4.9990$. This function is shown in Fig.~\ref{ph2}. 

\begin{figure}[t]
\begin{center}
\begin{minipage}[h]{0.47\linewidth}
\centering
\includegraphics[width=1\linewidth]{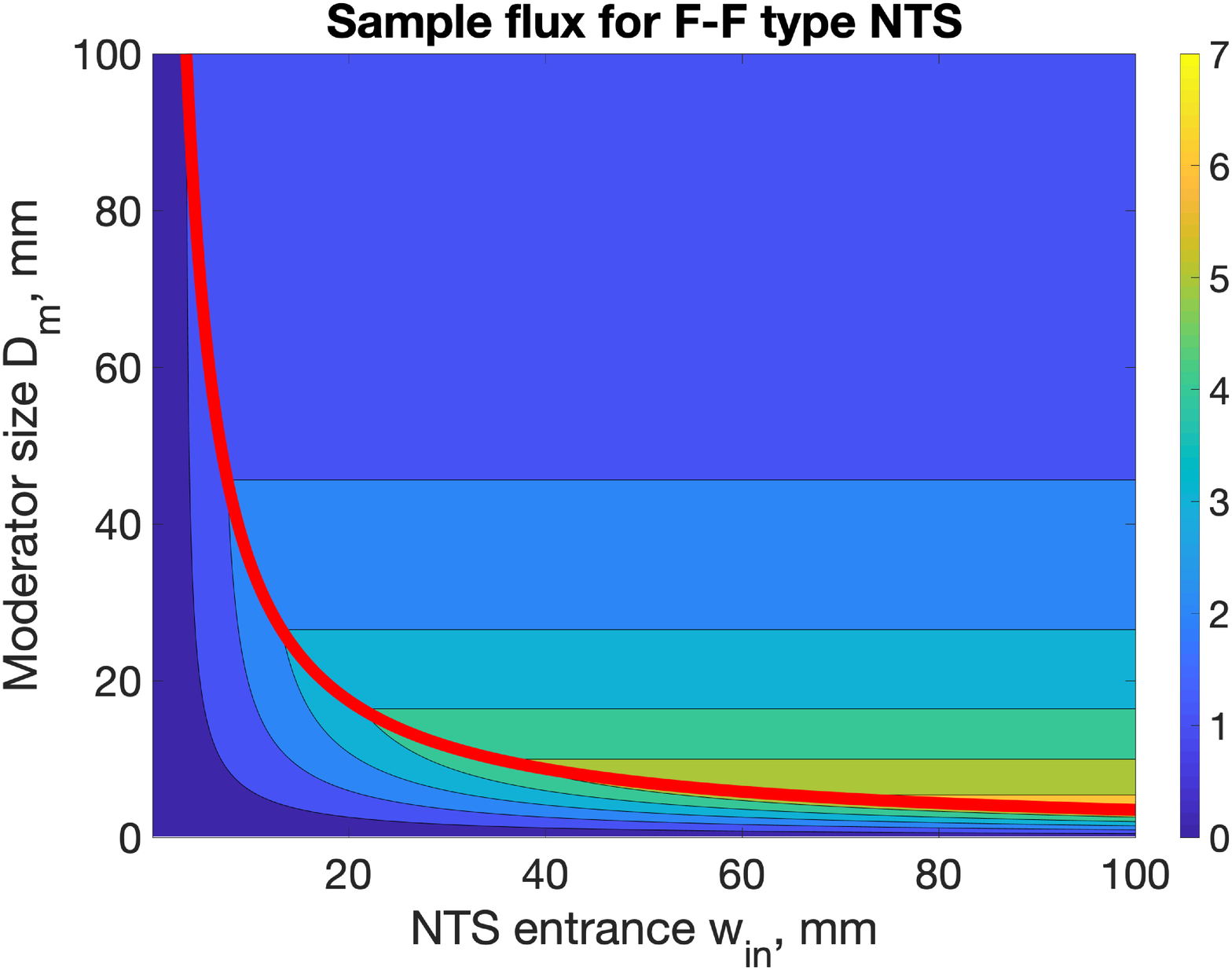} \\ (a)
\end{minipage}
\begin{minipage}[h]{0.47\linewidth}
\centering
\includegraphics[width=1\linewidth]{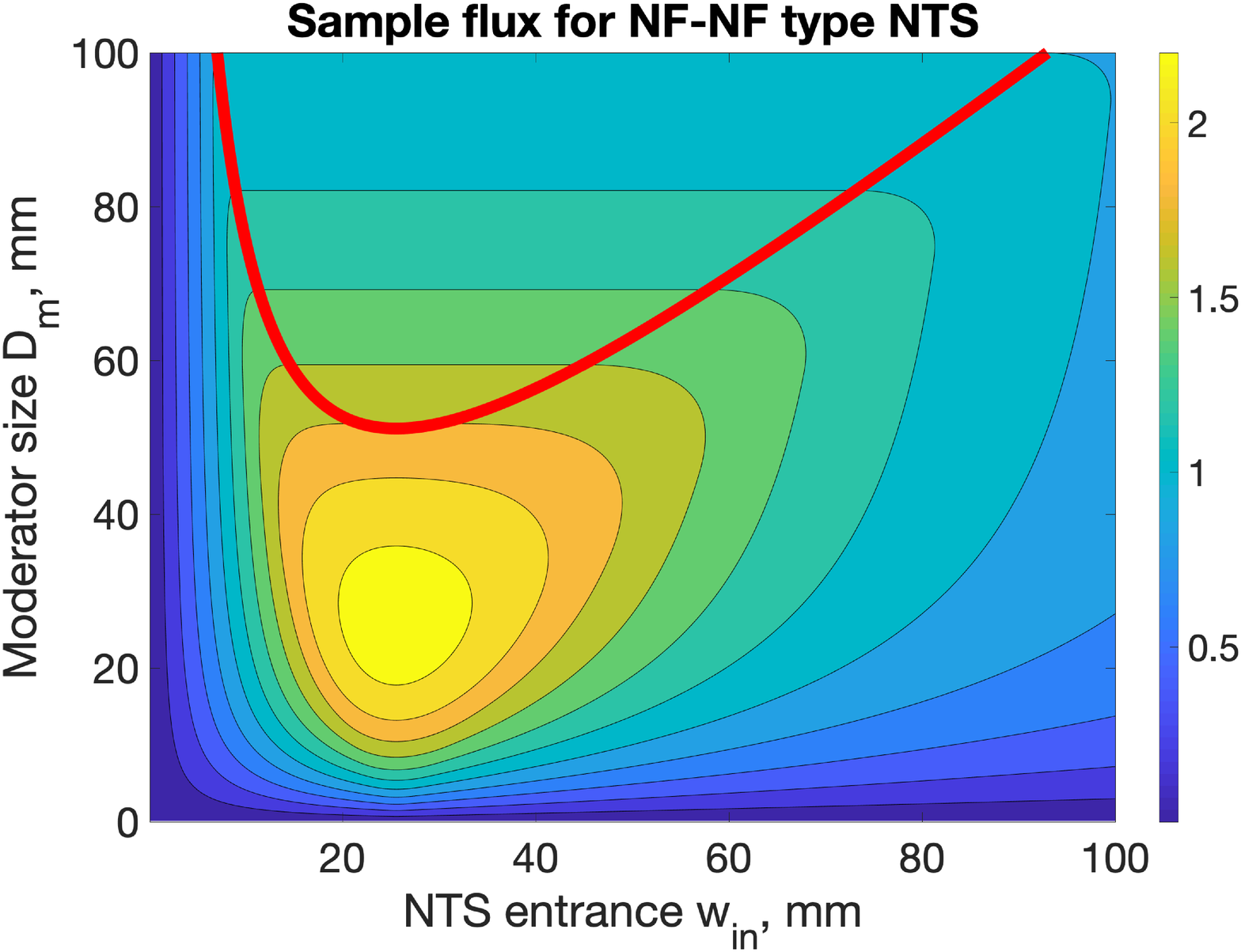} \\ (b)

\end{minipage}
\caption{Sample flux maps in case of employing para-H$_2$ moderator for F--F and NF-NF neutron transport systems. COFSIs are shown in red. Flux scales for (a) and (b) are different to keep a high colour contrast. Sample flux is normalized on the value when $B=1$.}
\label{parah2}
\end{center}
\end{figure}

Now let us consider the same instrument as described in Sec.~\ref{sample_flux_anal} but employing the above-mentioned para-H$_2$ moderator. To account for its brilliance we multiply the sample flux maps obtained above (Fig.~\ref{Sample_bril_map}) by function in Eq.~(\ref{bril_func}). Obtained sample flux maps are shown in Fig.~\ref{parah2}.

\begin{figure}[h!]
\begin{center}
\begin{minipage}[h]{0.45\linewidth}
\centering
\includegraphics[width=1\linewidth]{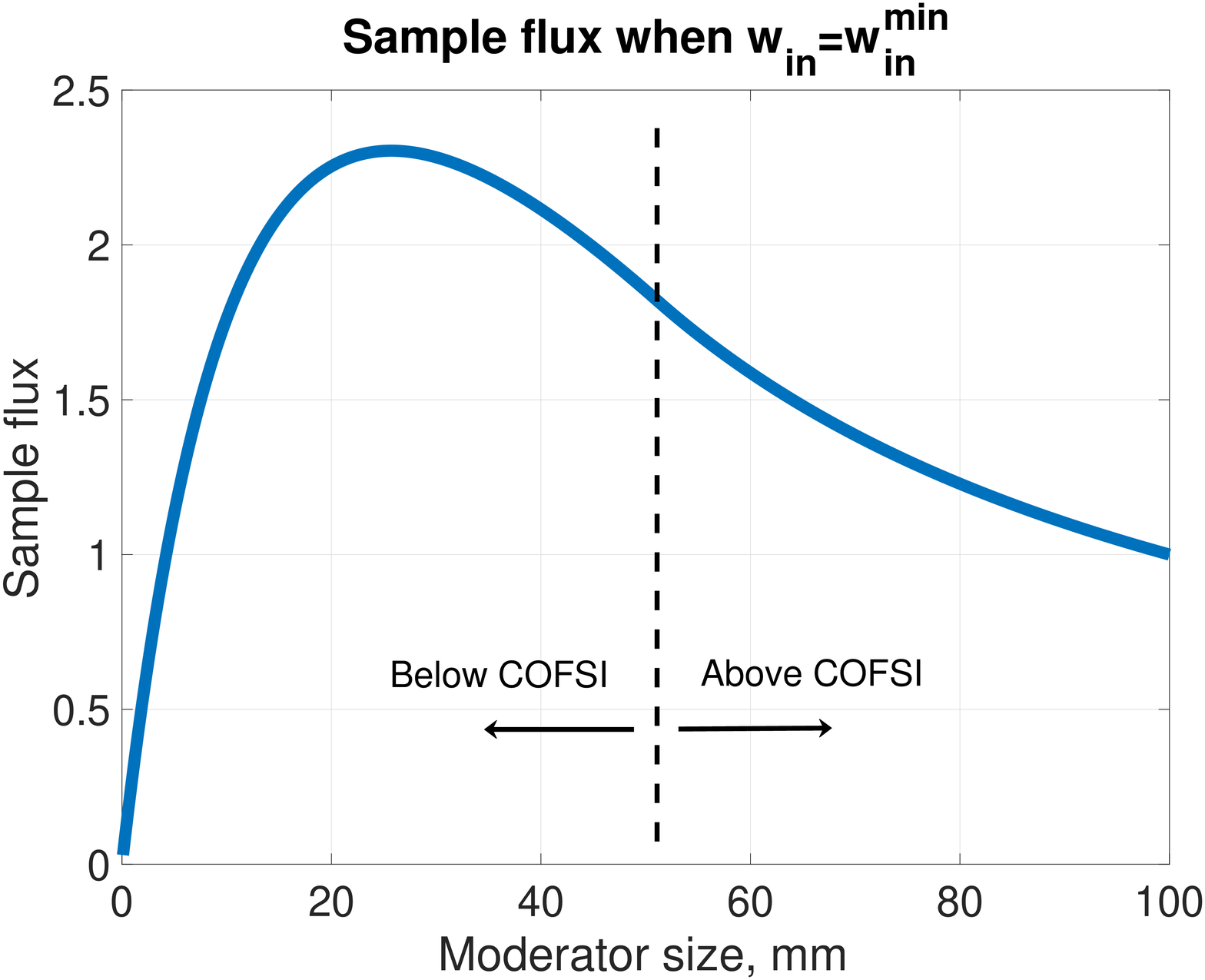}
\end{minipage}

\caption{Cross-section of sample flux map shown in Fig.~\ref{parah2}b.}
\label{cross}
\end{center}
\end{figure}

While all pairs $(w_{in}, D_m)$ belonging to the COFSI (shown in red) are still equivalent from the illumination point of view as they provide full sample illumination and minimal over-illumination, the sample flux is however different for each pair. While 100 mm moderator provides sample flux $\Phi_s=1$, smaller moderator allows to achieve even higher flux at the sample. In case of F--F type NTS it is possible to achieve very large sample flux of $\Phi_s=6.5$ for this particular instrument. This result highlights the importance of potential future search for NTSs with PS focusing entrances.

In case of NF--NF type NTS COFSI has a minimum at $w_{in}^{min}=25.5$~mm, $D_m^{min}=51$~mm, where sample flux reaches a value of $\Phi_s=1.8$.    Competing trends of sample under-illumination and increasing moderator brilliance provide maximal sample flux somewhere directly below the COFSI minimum as can be seen in Fig.~\ref{cross}. For this particular instrument maximal sample flux is $\Phi_s=2.3$, a gain of 1.28 over one reached with full sample illumination.

Since maximal sample flux is reached below the COFSI, it means the sample is under-illuminated. Then PS volume $V_s$ is not filled homogeneously, that usually results in an irregular beam divergence profile with multiple dips and peaks (see details in Sec.~\ref{sample_flux_anal} before Eq.~(\ref{inhomo}), also see \cite{mattauch2017}, Fig. 1). For low-resolution instruments, which integrate flux over large beam divergence, this “gothic”-like structure is not problematic. For high-resolution instruments, when $\alpha_s$ is comparable to peaks (dips) width and the angular precision of optical elements positioning, it could prove to be disastrous. The choice whether to work with the under-illuminated sample should be carefully considered in each particular case. Alternative “safe” option is to choose moderator size equal to COFSI minimum, which provides full sample illumination with slightly reduced sample flux.



%% file: two_modes.tex
\section{Practical applications of the developed method}
\label{practice}

\subsection{Case of an instrument with variable parameters}

The optimization method described in this paper allows to choose the moderator size basing on predefined instrument parameters, however quite often instruments operate in several modes with varied parameters.

Consider, for example, small-angle scattering instrument (SANS) performing an experiment over a wide range of the momentum transfer. Variable collimation base must be used to cover the whole required range, meaning that angular resolution is changed during the experiment. Other parameters like $d_s$ and $n=1.5$ (first collimation slit is twice as large as the sample) are constant and NF--NF type NTS is used.

The instrument optics and moderator size can be optimized only for one specific value $\alpha_s^*$. We define all other possible resolutions $\alpha_s = k \alpha_s^*$, which corresponds to the required PS volume $V_s = 2d_s\alpha_s$. 

\begin{figure}[h!]
\begin{center}
\begin{minipage}[h]{0.99\linewidth}
\centering
\includegraphics[width=1\linewidth]{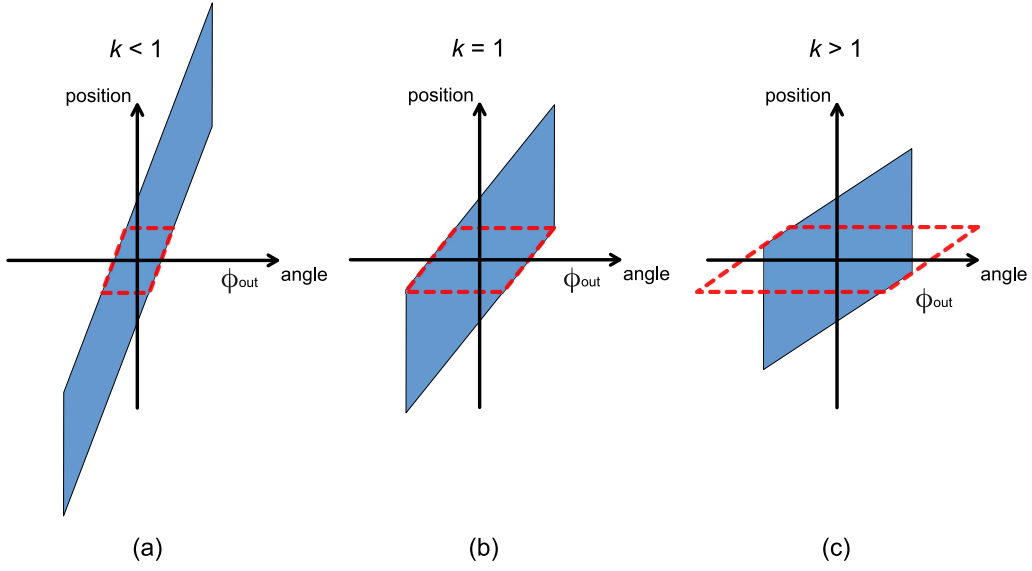}
\end{minipage}

\caption{PS representation of the neutron beam at the sample (shown in blue) and instrument requirements (shown with red dashed line) for different values $k$.}
\label{sans}
\end{center}
\end{figure}

Fig.~\ref{sans} shows PS representation of neutron beam at the sample position. PS volume $V_{out}$ (shown in blue) is optimized for optimal (i.e. minimal flux of undesirable neutron, see Sec.~\ref{samp_illum} for details) and full sample illumination ($k=1$). This optimization graphically corresponds to equal slopes of $V_{out}$ and $V_s$ (shown with red dashed line) sides. Since different collimation bases are used for $k\ne1$, the shape of $V_{out}$ is also different: its skewness directly depends on the distance between the optics exit and sample. Note that slopes of both volumes sides are equal in all cases, since they both depend linearly on $k$.

If $k<1$, then $V_s$ is fully filled (Fig.~\ref{sans}a): 

\begin{equation}
    V_s \cap V_{out} = 2 k d_s \alpha_s^*.
\end{equation}

Indeed, the sample is fully illuminated, however not optimally since too much of excessive PS volume $V_{out}$ is presented at the sample position, meaning relatively high flux of undesirable neutrons. For optimal sample illumination divergence $\phi_{out}$ should have been smaller by a factor $k$.

If $k>1$, then $V_s$ is not fully filled (Fig.~\ref{sans}c). Using that $n=1.5$ and $\phi_{out} = n\alpha_s$ as in Eq.~(\ref{1st_req}), we obtain:

\begin{equation}
    V_s \cap V_{out} =  \begin{cases}
            3d_s\alpha_s^* - d_s\alpha_s^* \dfrac{(3 - k)^2}{4}, & 1<k\le3 \\
        3 d_s \alpha_s^*, & k>3 .

    \end{cases}
\end{equation}
Thus, in this case sample is under-illuminated and effectively the instrument operates with better than required resolution.

Sample flux can be calculated according to Eq.~(\ref{brilliance}) as

\begin{equation}
    \Phi_s = b_{out} \times (V_s \cap V_{out}),
\end{equation}
where $b_{out}$ is the brilliance of the delivered neutron beam. This equation can be rewritten as

\begin{equation}
    \Phi_s = 
    \begin{cases}
    2 B k d_s \alpha_s^*, & k \le 1 \\
    3 B d_s\alpha_s^* - B d_s\alpha_s^* \dfrac{(3 - k)^2}{4}, & 1<k\le3 \\
    3 B d_s \alpha_s^*, & k>3 ,
    \end{cases}
\end{equation}
where $B=b_{out}$ is the moderator brilliance.

\begin{figure}[h]
\begin{center}
\begin{minipage}[h]{0.49\linewidth}
\centering
\includegraphics[width=1\linewidth]{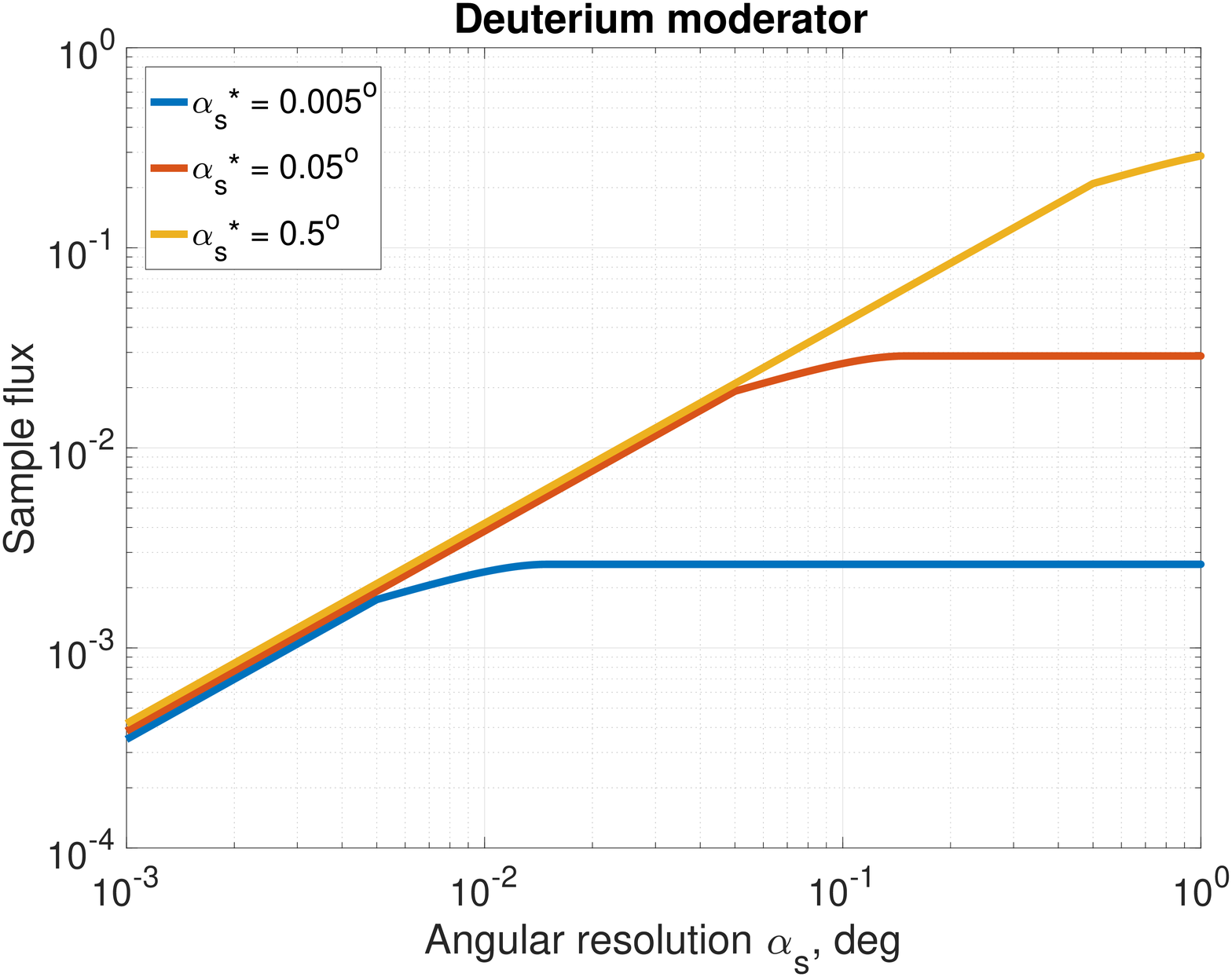} \\ (a)
\end{minipage}
\begin{minipage}[h]{0.49\linewidth}
\centering
\includegraphics[width=1\linewidth]{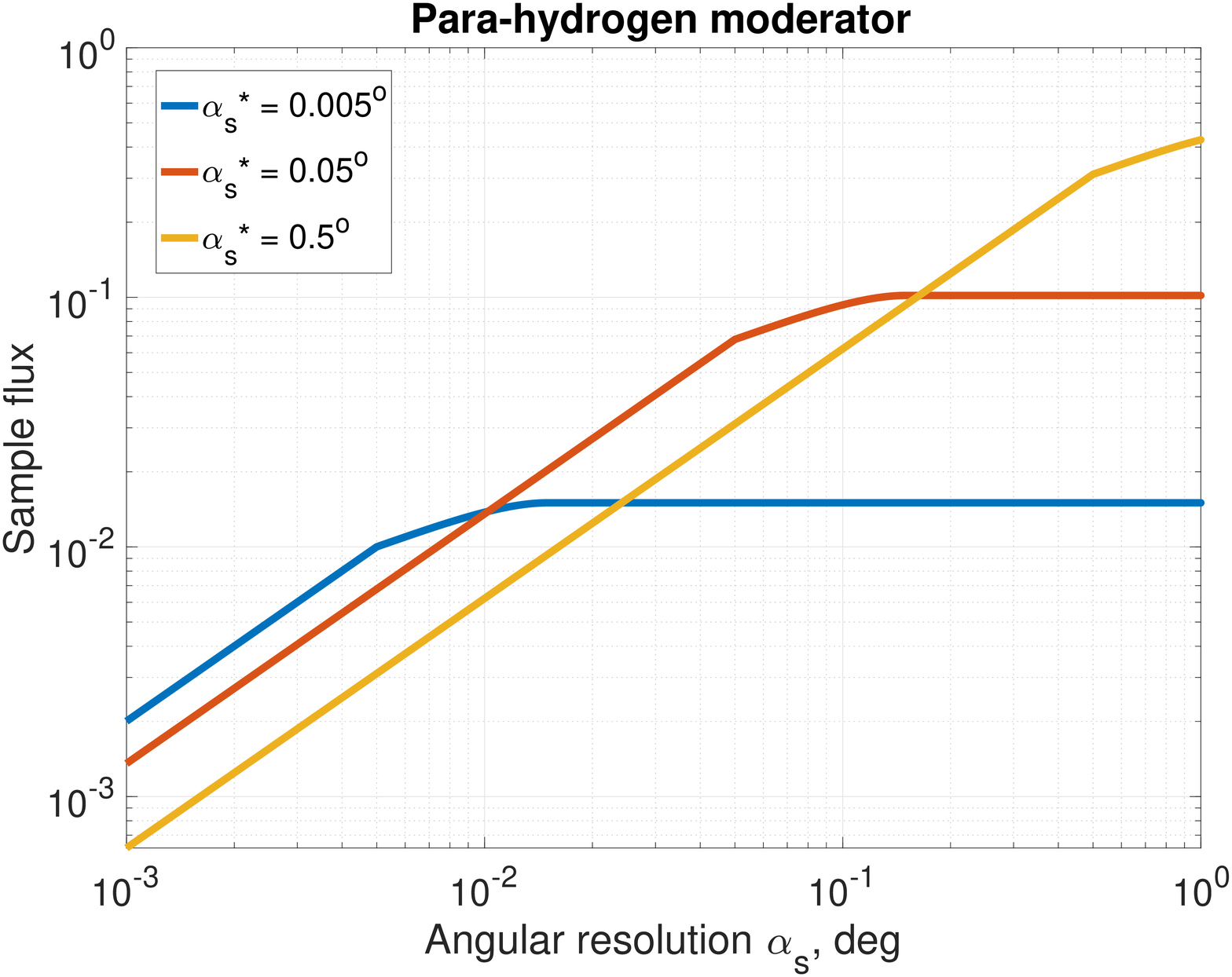}\\ (b)
\end{minipage} 

\caption{Sample flux for different angular resolutions of the SANS instrument in the case of using (a) deuterium moderator and (b) para-hydrogen moderator. Inclined lines in panel (a) are slightly offset vertically for better presentation.}
\label{sans_opt}
\end{center}
\end{figure}

Consider an instrument with parameters $d_s = 10$~mm, $L_{in} = 2000$~mm, $n=1.5$ and varied angular resolution $0.001^o \le \alpha_s \le 1^o$. The NTS and moderator can be optimized only for one chosen resolution $\alpha_s^*$. Results of sample flux calculations for different angular resolutions $\alpha_s^*$ are presented in Fig.~\ref{sans_opt}.

Let us again consider two types of moderators employed for the illumination of NTS. In case of large liquid deuterium moderator with size-independent brilliance (Fig.~\ref{sans_opt}a) it is always beneficial in terms of sample flux to optimize the NTS for the largest $\alpha_s^*$, i.e. for the coarse resolution, thus achieving maximal possible flux at the sample. Then for finer resolution, that corresponds to $k < 1$, the sample flux is reduced proportionally to $k$, as expected for a tightened resolution.

In case of para-H$_2$ moderator brilliance $B$ depends on its size. If the instrument was optimized for a specific $\alpha_s^*$, then the optimal moderator size is determined using Eqs.~(\ref{w_min_double_slit},\ref{D_min}). Corresponding moderator brilliance can be calculated using Eq.~(\ref{bril_func}). 

If the instrument employs such para-H$_2$ moderator, then it may be beneficial to optimize the NTS and the moderator size for small $\alpha_s^*$ (see Fig.~\ref{sans_opt}b). In the region of small $\alpha_s$ the optimization for small $\alpha_s^*$ (blue line) provides the sample flux gain of about 4 compared to the optimization for large $\alpha_s^*$ (yellow line). However, in the region of large $\alpha_s$ we get a loss of about 25. Hence, optimal solution for large $\alpha_s^*$ leads to losses for small $\alpha_s$ mode and vice versa. Optimization of the NTS and moderator for intermediate $\alpha_s^*$ (red line) can be seen as an option for compromise.

In a similar way it is possible to analyse other instruments performing experiments with variable $\alpha_s$ or $d_s$, e.g. reflectometers performing $\theta$--$2\theta$ scans.

%% file: TOF.tex
\newpage
\subsection{Case of time-of-flight instrument}

Many neutron instruments operate in time-of-flight mode, measuring scattering of neutrons with different wavelengths in the neutron pulse almost simultaneously. However, the geometry of NTS remains the same during the pulse and since the critical angle of reflection of the guide walls' coating is proportional to wavelength, the beam divergence $\phi_{out}$ at the NTS exit changes during the experiment.

According to Eq.~(\ref{1st_req}) optimal and full sample illumination can be achieved only for unique value of $\phi_{out}^*$ and, indeed, only for unique neutron wavelength $\lambda^*$, that in turn determines the optimal moderator size $D_{opt}$. For other wavelengths $\lambda = p \lambda^*$, $\phi_{out} = p \phi_{out}^*$ and the chosen moderator size is not optimal.

For simplicity, we consider the case of NF--NF type NTS and instrument with Soller collimator ($n=1$), which corresponds to the rectangular phase space volume $V_s$. Expressions for other types of NTSs or for $n>1$ can be derived in a similar way, however are more cumbersome. 

\begin{figure}[h!]
\begin{center}
\begin{minipage}[h]{0.5\linewidth}
\centering
\includegraphics[width=1\linewidth]{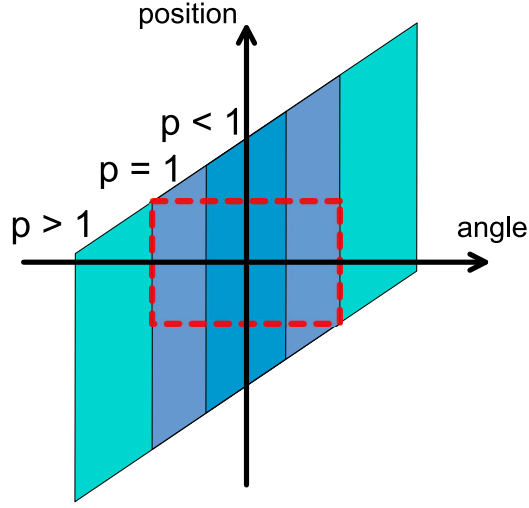}
\end{minipage}

\caption{PS volume $V_{out}$ at the sample position. Different shades of blue correspond to different values of $p$. Shown with red dashed line is $V_s$.}
\label{tof_nf_volume}
\end{center}
\end{figure}

The PS volume $V_{out}$ of the beam at the NTS exit is  different from the optimal one by factor $p$:

\begin{equation}
    V_{out} = p V_{out}^*.
    \label{exit_volume}
\end{equation}
The shape of $V_{out}$ at the sample position is shown in Fig.~\ref{tof_nf_volume}.

If $p>1$, then PS volume $V_s$ required by the instrument is fully inscribed in $V_{out}$. Otherwise it is filled only partially, meaning that PS volume $V_s \cap V_{out}$ actually available to the instrument is reduced:

\begin{equation}
V_s \cap V_{out} =
    \begin{cases}
 V_s , & p \ge 1 \\
 p V_s , & p < 1.
\end{cases}
\label{sample_volume_calc}
\end{equation}

Excessive PS volume $V_{out} - V_s \cap V_{out}$ at the sample position may provide undesirable background. The ratio of “useful” to “useless” neutrons is independent from $p$ for $p < 1$ and decreases inversely proportional to $p$ for $p\ge1$.

\begin{figure}[h]
\begin{center}
\begin{minipage}[h]{0.75\linewidth}
\centering
\includegraphics[width=1\linewidth]{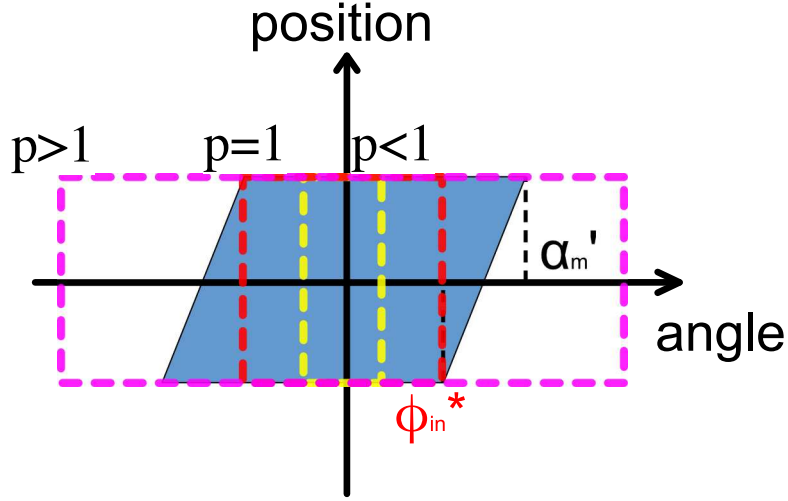}
\end{minipage}

\caption{PS volume $V_{in}$ (shown with dashed line) at the NTS entrance. Different colors of dashed lines correspond to different values of $p$. Shown in blue is $V_m$.}
\label{tof_entrance}
\end{center}
\end{figure}

Consider now the situation at NTS entrance. The beam divergence $\phi_{in} = p \phi_{in}^*$, that can be accepted by NTS, also changes proportionally to neutron wavelength, meaning that potentially acceptable PS volume is $V_{in} = p V_{in}^*$, where $V_{in}^*$~--- PS volume accepted by the NTS when $\lambda=\lambda^*$ (see Fig.~\ref{tof_entrance} and compare to Fig.~\ref{small_d_source_cap}).

If $p<1$ the NTS entrance is over-illuminated and brilliance $b_{out}$ at the sample position is equal to moderator brilliance $B$.

If $p>1$ the NTS entrance is under-illuminated and according to Eq.~(\ref{inhomo}) brilliance $b_{out}$ at the sample position is reduced compared to $B$:

\begin{equation}
b_{out} = B \dfrac{V_{in} \cap V_m}{V_{in}} = 
    \begin{cases}
 \dfrac{V_m}{V_{in}} B, & p > \dfrac{\alpha_m'}{\phi_{in}^*}, \\
 B  - B \dfrac{w_{in}(\phi_{in}^*)^2}{V_{in}}\dfrac{(p-1)^2}{\alpha_m' - \phi_{in}^*}, & 1 < p \le \dfrac{\alpha_m'}{\phi_{in}^*}, \\
 B, & p \le 1.
\end{cases}
\label{bril_tof}
\end{equation}

Taking into account Eqs.~(\ref{div_source1})--(\ref{moderator_volume}) and (\ref{div_equ}) we can rewrite Eq.~(\ref{bril_tof}) as:

\begin{equation}
b_{out} =
    \begin{cases}
 B \dfrac{D_{opt}}{ p (D_{opt} - w_{in})}, & p \ge \dfrac{D_{opt} + w_{in}}{D_{opt} - w_{in}}, \\
 B \left(1 - \dfrac{(p-1)^2 (D_{opt} - w_{in})}{4 p w_{in}}\right), & 1 < p \le \dfrac{D_{opt} + w_{in}}{D_{opt} - w_{in}}, \\
 B, & p \le 1.
\end{cases}
\label{brilliance_non_cofsi}
\end{equation}

Sample flux for any given $\lambda$ can be calculated according to Eq.~(\ref{brilliance}) as

\begin{equation}
    \Phi_s = b_{out} \times (V_s\cap V_{out}).
    \label{tof_sample_fluxx}
\end{equation}

Substituting expressions ~(\ref{sample_volume_calc}) and (\ref{brilliance_non_cofsi}) in Eq.~(\ref{tof_sample_fluxx}) we obtain

\begin{equation}
    \Phi_s = \begin{cases}
         B V_s \dfrac{D_{opt}}{ p (D_{opt} - w_{in})}, & p \ge \dfrac{D_{opt} + w_{in}}{D_{opt} - w_{in}} , \\
 B V_s \left(1 - \dfrac{(p-1)^2 (D_{opt} - w_{in})}{4p w_{in}}\right), & 1 \le p < \dfrac{D_{opt} + w_{in}}{D_{opt} - w_{in}} , \\
 BV_sp, & p < 1.
    \end{cases}
    \label{tof_flux_eq}
\end{equation}

\begin{figure}[h!]
\begin{center}
\begin{minipage}[h]{0.5\linewidth}
\centering
\includegraphics[width=1\linewidth]{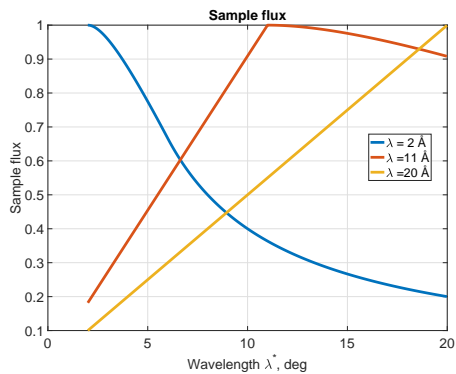}
\end{minipage}

\caption{Sample flux as a function of neutron wavelength for the NTS and moderator size optimized for particular neutron wavelengths.}
\label{tof_flux}
\end{center}
\end{figure}

Consider an instrument working in the range $\lambda=2-20$~\AA{}. The NTS and the moderator size are optimized for one specific $\lambda^*$. Fig.~\ref{tof_flux} demonstrates the $\lambda$-dependence of sample flux. Sample flux rises linearly up to this wavelength and then falls afterwards. For $\lambda^*$ on the border of the $\lambda$ range only half of this curve is presented. There are two ways to proceed to choose $\lambda^*$ for which to optimize the NTS and moderator size. 

\begin{enumerate}
    \item If some particular neutron wavelength is of most interest, then it should be chosen as $\lambda^*$. In this case sample flux reaches maximum exactly when $\lambda=\lambda^*$.
    \item Alternatively, one can aim for maximal integrated sample flux in single neutron pulse. From three options shown in Fig.~\ref{tof_flux} the best one would be $\lambda^*=11$~\AA{}.
\end{enumerate}

In practice, additional factors should be additionally taken into account in Eq.~(\ref{tof_flux_eq}). First, neutron beam brilliance depends on wavelength because of wavelength dependence of moderator spectrum and NTS transmission. Exact expression for this dependence is different for different instruments. Second, if the constant relative resolution $\dfrac{\Delta\lambda}{\lambda}$ is employed, then flux at longer wavelengths is relatively higher than  at shorter wavelengths. Finally, the sample scattering power may depend on $\lambda$ as well. Taken all together, these factors will change the shape of curves shown in Fig.~\ref{tof_flux}, however the general conclusion about optimization options will not be affected.

The considerations in this section are related to the choice of the optimal wavelength for a time-of-flight instrument. We have provided a way to choose the wavelength to optimize the instrument for in case of time-of-flight technique is employed. Similar arguments may be applied to monochromatic instruments, using several selected wavelengths for different tasks.

%% file: two_instruments.tex
\subsection{Case of two instruments sharing the same moderator}

Let us consider two instruments both using NF--NF type NTSs and sharing the same para-H$_2$ moderator. Moderator size should be chosen to provide the best performance for both instruments. For simplicity, both instruments have the same parameters: $d_s=10$~mm, $L_{in}=2000$~mm and $L_{out}=500$~mm. The only difference are requirements for angular resolution: one instrument is a high-resolution (HR) one with $\alpha_s=0.1$\textdegree{}, while another is a low-resolution (LR) one with $\alpha_s=0.5$\textdegree. 

\begin{figure}[h!]
\begin{center}
\begin{minipage}[h]{0.47\linewidth} 
\centering
\includegraphics[width=1\linewidth]{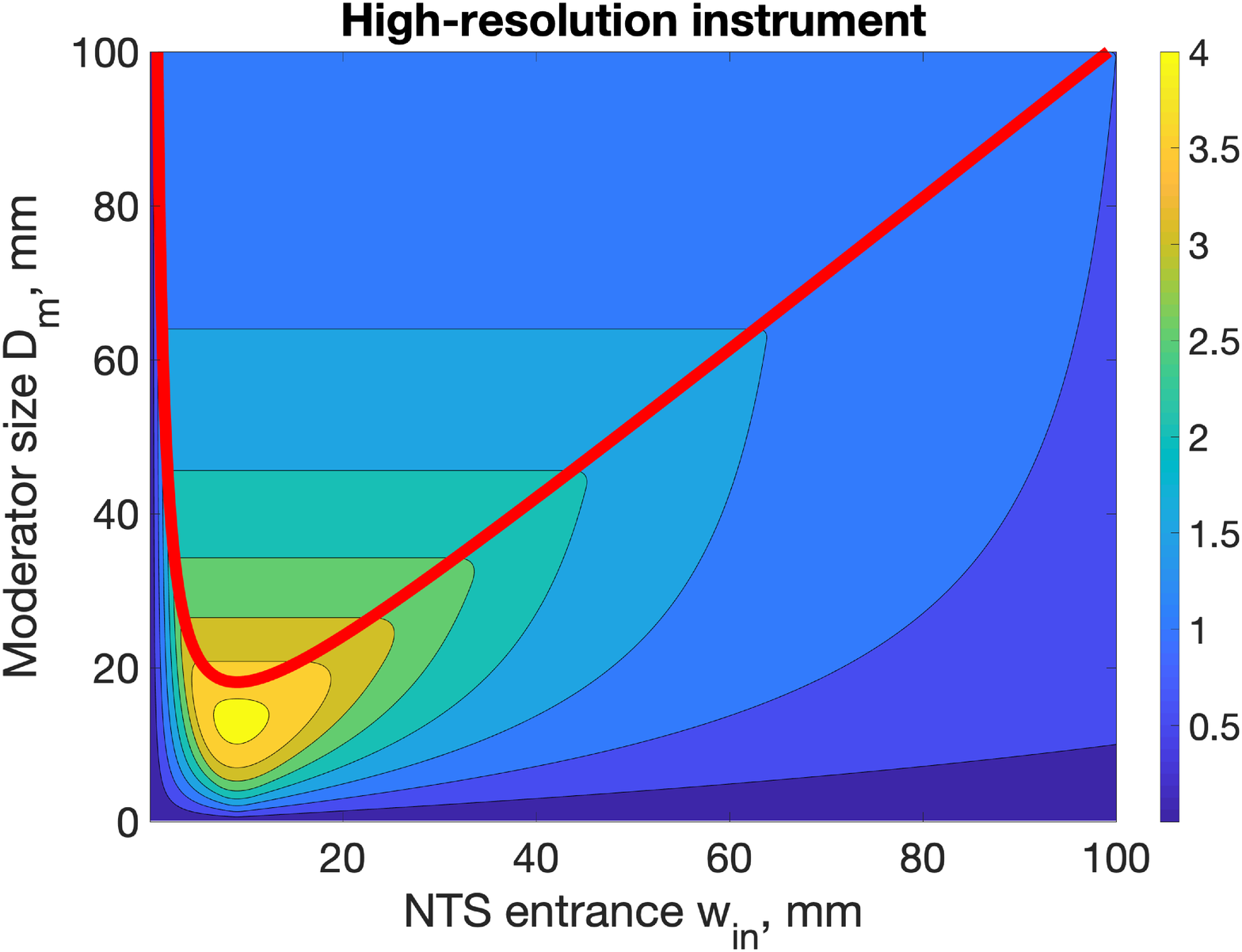} \\ (a)

\end{minipage}
\begin{minipage}[h]{0.47\linewidth} 
\centering
\includegraphics[width=1\linewidth]{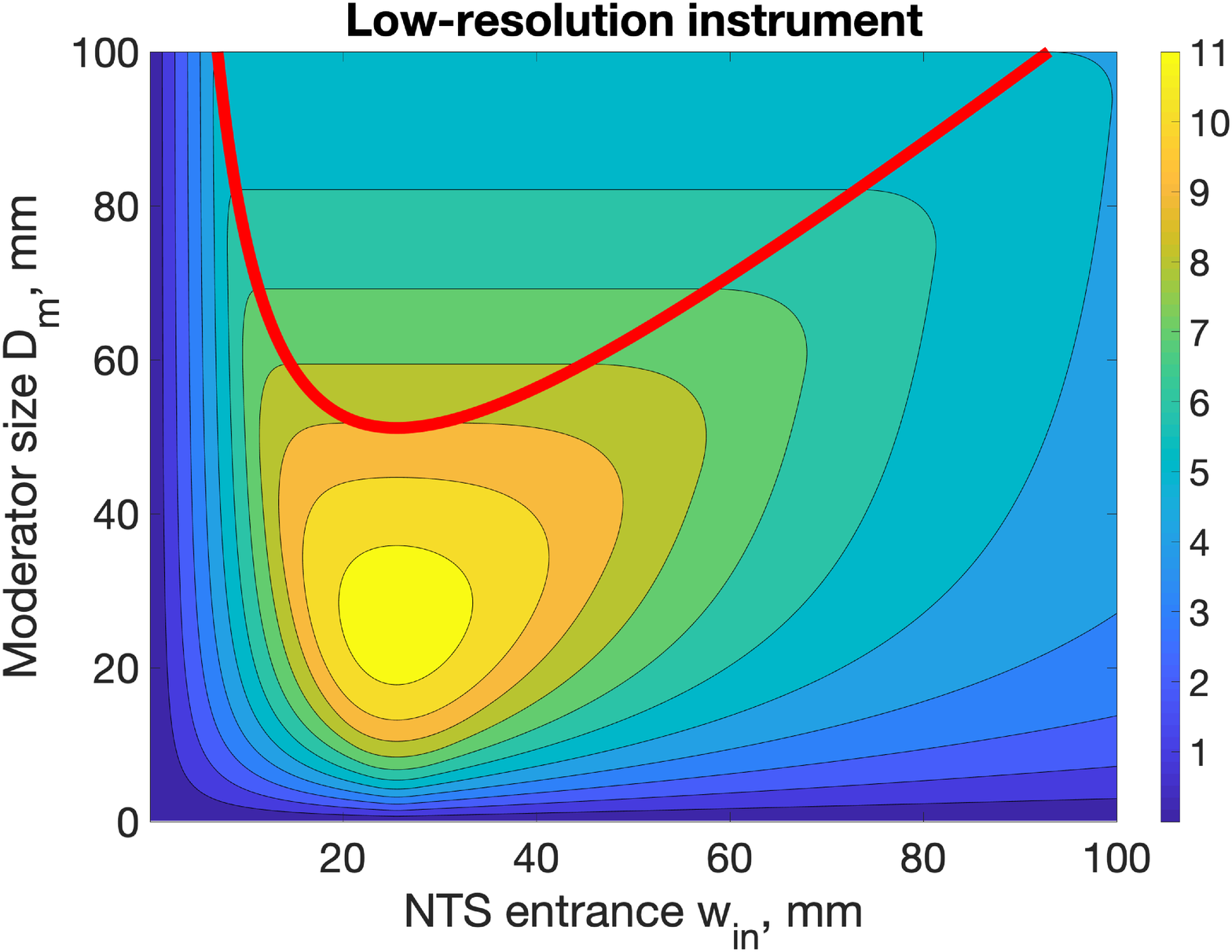} \\ (b)

\end{minipage}
\caption{Sample flux maps in case of using para-H$_2$ moderator for two instruments. COFSIs are shown in red. Flux scales for (a) and (b) are different to keep high colour contrast.}
\label{modes}
\end{center}
\end{figure}

Sample flux maps corresponding to both these instruments are shown in Figs.~\ref{modes}a,b, respectively. Different optimal sizes of moderator correspond to these instruments (Table~\ref{opt_mod_sizes}). For each of them there are two optimal solutions are provided: one for the case of optimal and full sample illumination (COFSI minimum in Fig.~\ref{modes}) and one for the case of under-illuminated sample with highest possible sample flux (yellow spot below COFSI in Fig.~\ref{modes}). Detailed comparison of these two options is given in Sec.~\ref{para-h2}. Note that despite a 5-times better resolution, the sample flux for high-resolution instrument (if optimised) is only 2.4--2.8 times less than sample flux for a low-resolution instrument, thanks to the increased brilliance of para-$H_2$ moderator.

\begin{table}[h]
\begin{center}
\begin{tabular}{|c|c|c|}
\hline
& HR instrument & LR instrument \\
\hline

\begin{tabular}{@{}c@{}}Optimally and fully \\ 
 illuminated sample \end{tabular} & \begin{tabular}{@{}c@{}} $D_{opt} = 18$~mm \\ 
 $\Phi_s = 3.8$ \end{tabular}& \begin{tabular}{@{}c@{}} $D_{opt} = 51$~mm \\ 
 $\Phi_s = 9$ \end{tabular}\\
 \hline
  \begin{tabular}{@{}c@{}}Under-illuminated sample; \\ 
 maximal sample flux \end{tabular} & \begin{tabular}{@{}c@{}} $D_{opt} = 13$~mm \\ 
 $\Phi_s = 4.1$ \end{tabular}& \begin{tabular}{@{}c@{}} $D_{opt} = 25$~mm \\ 
 $\Phi_s = 11.5$ \end{tabular} \\
\hline
\end{tabular}
\caption{Optimal moderator sizes and corresponding sample fluxes for HR and LR instruments.}
\label{opt_mod_sizes}
\end{center}
\end{table}

There are two possibilities for the choice of the moderator size, suitable for both instruments:

\begin{enumerate}
    \item To optimize the moderator size for the best performance of low-resolution instrument. Moderator size $D_m=51$~mm is optimal if homogeneous sample illumination is required. For high-resolution instrument this choice means the sample flux of $\Phi_s^{HR}=1.8$. However, this value is significantly lower than the maximally achievable flux of 4.1.
    \item To optimize the moderator size for the best performance of high-resolution instrument. The moderator size $D_m=18$~mm is optimal and allows for homogeneous sample illumination. With such moderator the sample flux at low-resolution instrument is $\Phi_s^{LR} = 11$, which is slightly less than the maximally achievable flux of 11.5. Note, that for such choice of moderator size, the low-resolution instrument operates with under-illuminated sample.
\end{enumerate}

Particularly, for this pair of instruments we would suggest to optimize the moderator size for the best performance of high-resolution instrument allowing to obtain high sample flux for both instruments, because low-resolution instrument probably can perform reasonably well also with under-illuminated sample.

Similar considerations can be applied to instrument suits including many versatile instruments.

%% file: outro.tex
\section{Conclusion}

We have developed a simple analytic method to find out optimal combinations of sizes of moderator and entrance of the neutron transport system (NTS), that provides the full illumination of the sample with minimum to none over-illumination (as well as minimum to none background along the NTS) for any neutron scattering instrument. Only the knowledge of basic instrument parameters~--- sample size, angular resolution, distances from the NTS to moderator and sample, is required for calculations. When employing the low dimensional para-hydrogen moderators this method allows to find out the unique optimal solution, which provides the maximum sample flux. 

One of the important advantages of this method is that extensive time-consuming Monte-Carlo simulations, usually employed to tackle such problems, are not required. The optimizations of moderator and NTS are effectively decoupled and the number of free parameters for the neutron optics optimization is reduced. Monte-Carlo analysis can be used as a complimentary technique taking into account the non-ideal neutron transport and phase space focusing properties of a particular NTS.

This method can be used at initial steps of neutron sources/instrument design, for upgrades of NTS in the case of fixed moderator size or during the exchange/upgrade of neutron moderators to adapt them to parameters of existing neutron guides delivering neutrons from moderator to sample.


We have also shown that by means of the phase space focusing (F--F) type NTS it is principally possible to make use of very small para-hydrogen moderators with significantly enhanced brilliance even for neutron scattering instruments with a large sample and coarse angular resolution.